\documentclass[aps,prd,twocolumn,superscriptaddress,10pt]{revtex4-2}
\usepackage[T1]{fontenc}
\usepackage{inputenc}
\usepackage{babel}
\usepackage{natbib}
\usepackage{hyperref}
\hypersetup{colorlinks=true, citecolor=blue, urlcolor=blue, linkcolor=blue}
\usepackage{graphicx}
\usepackage{array}
\usepackage{multirow}
\usepackage{amsmath,amssymb}
\usepackage{mathrsfs}
\usepackage{physics}
\usepackage{longtable, array, booktabs}
\usepackage[table]{xcolor}
\usepackage{orcidlink}
\usepackage{float}

\begin{document}
\title{An efficient Wavelet-Based Hamiltonian Formulation of Quantum Field Theories using Flow-Equations}
\author{Mrinmoy Basak, \orcidlink{0000-0001-8965-0824}}
\email[Electronic Address: ]{mrinmoy.263009@gmail.com}
\author{Debsubhra Chakraborty, \orcidlink{0000-0001-5815-4182}}
\email[Electronic Address: ]{debsubhra.chakraborty@tifr.res.in}
\author{Nilmani Mathur, \orcidlink{0000-0003-2422-7317}}
\email[Electronic Address: ]{nilmani@theory.tifr.res.in}
\affiliation{Deparment of Theoretical Physics, Tata Institute of Fundamental Research, Mumbai, 400005, India}
\begin{abstract}
We propose an effective Hamiltonian formulation of quantum field theories using a Daubechies wavelet basis in position space. Combined with flow-equation methods of the similarity renormalization group (SRG), this approach provides an efficient framework for analyzing quantum field theories by reducing the dimensionality of the Hamiltonian and systematically decoupling degrees of freedom across scales. As an application, the free scalar field theory has been reformulated within this framework to calculate the low-lying energy spectrum of the theory. These basis elements are known to transform the free scalar field theory into a theory of coupled localized oscillators, each of which is labeled by a location and a resolution index. In this representation, the Hamiltonian is naturally organized into fixed-resolution blocks, alongside blocks associated with the interactions between different resolutions. To decouple the different resolution modes and obtain a block diagonalized Hamiltonian with each block associated with a fixed resolution, the flow equation approach of SRG is applied. Finally, we demonstrate that with  increasing resolution, the low-energy spectrum can be extracted from the effective lowest-resolution block of the Hamiltonian, leading to a significant reduction in computational cost.
\end{abstract}
\maketitle
\section{Introduction}
\label{sec:Introduction}
The origin of Daubechies wavelets can be traced back to 1909, when Haar introduced a system of piecewise constant functions on $\left[0,1\right]$~\cite{Haar1910wavelets}. The primary motivation was to construct an compactly supported orthonormal basis, in contrast to the Fourier basis, in order to better represent localized features such as sudden bumps in a function. Though the Haar system of orthogonal functions was identified as an efficient tool to address problems in approximation and martingale theory~\cite{Getoor2009martingle}, its discontinuity restricted its applicability to smooth functions. During the late 1980s, Ingrid Daubechies generalized the idea of compactly supported wavelets by constructing a family of orthonormal bases with higher regularity and improved approximation properties for smooth functions~\cite{https://doi.org/10.1002/cpa.3160410705}. The basis functions are discrete and characterize by the location and the length scale (or resolution). During the same time, the theory of multiresolution analysis was introduced by Mallat~\cite{10.1109/34.192463}, which provides a hierarchical decomposition of a signal into nested approximation spaces and corresponding detail spaces at different scales. Daubechies wavelets were constructed within the multiresolution analysis framework, providing compactly supported orthonormal bases suitable for multiscale signal decomposition. 

On the other hand, in the context of quantum field theory, ideas closely related to wavelets first appeared in the work of Kenneth G. Wilson~\cite{PhysRev.140.B445}. 
In this work, he proposed a function, which is either localized in the position space and delocalized in the momentum space and vice-versa. However, a full rigorous mathematical construction of those functions was not provided. He also attempted to construct a function which would have simultaneous localization in position and momentum space, but it was not realized at that time. After the introduction of compactly supported wavelet bases by Daubechies (1994)~\cite{https://doi.org/10.1002/cpa.3160410705}, Wilson proposed their potential use in multiscale Hamiltonian formulations and encouraged further development of these ideas~\cite{PhysRevD.49.6720}. 

Over the next three decades, several authors worked towards the development of wavelet-based quantum field theoretical frameworks~\cite{10.1006/acha.1994.1016,10.1143/PTP.94.1135,10.1007/s00601-018-1357-z,michlin2017using,10.1143/PTP.94.1135,BEST2000848,PhysRevLett.116.140403,10.1007/JHEP06,PhysRevD.106.036025,PhysRevD.108.125008,best1994variationaldescriptionstatisticalfield,HALLIDAY1995414,10.1063/1.1543582,10.1088/1751-8121/ad5503,PhysRevD.87.116011,10.1051/epjconf/201817511002,10.3842/SIGMA.2007.105,albeverio2010remarkgaugeinvariancewaveletbased,10.1134/S1063778818060029,PhysRevD.88.025015,10.1007/s11182-013-9940-8,10.1007/s10773-015-2913-7,polyzou2020lightfrontquantummechanicsquantum,PhysRevD.107.036015,PhysRevD.111.096024,74c7-y1gp,PhysRevD.101.096004,10.1007/s00601-003-0008-0,basak2025multiresolutionanalysisquantumtheories}. During this period, these ideas were further developed in the context of Hamiltonian formulations by W. N. Polyzou and collaborators. In their 2013 work, the Daubechies wavelet basis was employed to develop a Hamiltonian formulation of the free scalar field theory, exhibiting sparse couplings between modes at the same and different scales~\cite{PhysRevD.87.116011}. 
To be noted that the creation and annihilation operators defined in Ref.~\cite{PhysRevD.87.116011} assume that a scaling mode can be constructed from a single scaling-field variable and its conjugate momentum.  However, such a definition is admissible only when the quadratic Hamiltonian is diagonal in the mode basis. On the contrary, in the Daubechies wavelets representation in a position space, the scaling modes remain coupled to one another, and therefore the above construction does not provide a consistent particle interpretation. Realizing this  inconsistency,  in the present work, we define a fully consistent scaling mode creation and annihilation operator to perform the second quantization of the free scalar field theory. A detailed demonstration of this subtle inconsistency is provided in Appendix \ref{appen:remarks_on_the_creation_and_annihilation_operator_construction}. 

{}

To numerically analyze a quantum field theory within this formalism, truncation in both the infrared and ultraviolet regions is required. However, the scaling and wavelet modes remain coupled even after truncation. As a result, construction of a Fock basis with both sets of modes leads to a rapid increment in the number of basis states and, consequently, in the size of the Hamiltonian. Hence, it is desirable to employ a systematic procedure to decouple the degrees of freedom associated with different scales. A natural framework for achieving such a decoupling is provided by the similarity renormalization group (SRG), which implements continuous unitary transformations of the Hamiltonian to a block diagonal form,  with each block associated with the different scale degrees of freedom~\cite{PhysRevD.95.094501,PhysRevD.111.096024}. As a result, the low-lying eigenvalues of the theory can be computed by representing the Hamiltonian matrix in the Fock basis states constructed using only the coarser-scaled modes.

The formulation of SRG was first introduced by Wegner~\cite{wegner1994flow} and independently by Glazek and Wilson~\cite{glazek1993renormalization,glazek1994perturbative}. In his 1994 paper~\cite{PhysRevD.49.6720}, Wilson  extended SRG as a method to decouple high- and low-energy degrees of freedom by transforming the Hamiltonian into a band-diagonal form, formulated in terms of light-front momentum variables. Michlin et al.~\cite{PhysRevD.95.094501} were the first to study of the flow-equation methods of SRG for a $1+1$-dimensional real scalar field theory in the instant form, using a wavelet-based representation. We previously extended this formulation to study a quantum field theory with quadratic interactions~\cite{PhysRevD.111.096024}. In the present work, we further develop the $1+1$-dimensional free scalar field theory in two ways: (i) by decoupling low- and high-resolution degrees of freedom using flow-equation methods up to three additional resolution levels (reaching resolution 4), and (ii) by defining creation and annihilation operators solely in terms of effective scaling variables, resulting in a significant reduction in the size of the Fock space basis.


The paper is organized as follows. In Sec.~\ref{sec:introduction_to_the_daubechies_wavelet_basis}, we introduce Daubechies wavelets and review properties relevant to quantum field theory. In Sec.~\ref{sec:The_Flow_equation_method_in_Daubechies_wavelet_basis}, we describe the application of the flow-equation method within the wavelet-based framework. The block-diagonalization of the 1+1-dimensional free scalar field theory using flow-equation methods is then presented in Sec.~\ref{sec:Effective_Hamiltonian_from_Flow_Equations}. In Sec.~\ref{sec:second_quantization}, we perform the second quantization of the free scalar field theory by defining creation and annihilation operators in terms of the effective scaling degrees of freedom and compute the low-lying eigenvalues by truncating the Hilbert space in the ultraviolet and infrared regions. The summary and concluding remarks are presented in Sec.~\ref{sec:conclusion_and_outlook}.

\section{Introduction to the Daubechies wavelet basis}
\label{sec:introduction_to_the_daubechies_wavelet_basis}
In this section, we introduce the essential components of the Daubechies wavelet basis to be used throughout the paper. For a more detailed study, the reader may refer to Refs.~\cite{kessler2003waveletnotes,daubechies1992ten}.

These basis functions are chosen because they are orthogonal and compactly supported, enabling the theory to be expressed in terms of local variables. This is analogous to lattice field theory, which is also formulated using local degrees of freedom. The multiresolution nature of these basis functions allows local operators to be represented in terms of variables at different scales, without explicitly incorporating all small-scale components.

The scaling-wavelet basis elements can be formed using a single function, $s(x)$, called the mother scaling function, defined through the following renormalization group equation,
\begin{eqnarray}
\label{eq:scaling_equation}
    s(x) =\sum_{l=0}^{2K-1}h_l \hat{D} \hat{T}^ls(x),
\end{eqnarray}
where, $\hat{T}$ and $\hat{D}$ are the unitary translation and unitary dyadic scale transformation operators respectively. $\hat{T}$ shifts the scaling function towards the right and $\hat{D}$ shrinks the function to $\frac{1}{2}$ of it's original value, while keeping the norm, $\int s(x)dx=1$, fixed,
\begin{eqnarray}
    \hat{D}s(x)=\sqrt{2}s(2x),\quad \hat{T}s(x)s(x-1).
\end{eqnarray}
$K$ is called the order of the scaling function, which determines the amount of smoothness and the support of the function. The coefficients $h_n$ are real numbers that can be determined by solving a set of three equations obtained by imposing the following conditions on the mother scaling function:
\begin{enumerate}
    \item The mother scaling function $s(x)$ and its integer translates $s(x-n)$ are orthonormal,
    \begin{eqnarray}
    \label{eq:orthonormality_of_the_mother_scfaling_function}
        \int s(x)s(x-n)\,dx = \delta_{0n},
    \end{eqnarray}
    which yields the relation
    \begin{eqnarray}
    \label{eq:h_n_1st_equation}
        \sum_{m=0}^{2K-1} h_m\, h_{m-2n} = \delta_{0n}.
    \end{eqnarray}

    \item The integral of the mother scaling function is normalized to unity,
    \begin{eqnarray}
        \int s(x)\, dx = 1,
    \end{eqnarray}
    which gives
    \begin{eqnarray}
    \label{eq:h_n_2nd_equation}
        \sum_{n=0}^{2K-1} h_n = \sqrt{2}.
    \end{eqnarray}

    \item Any polynomial of degree $m$ ($0 \le m < K$) can be expressed in terms of the mother scaling function and its translates,
    \begin{eqnarray}
        x^m = \sum_{n=-\infty}^{\infty} c_n\, s(x-n),
    \end{eqnarray}
    which leads to
    \begin{gather}
        \sum_{n=0}^{2K-1} n^{m} g_n
        = \sum_{n=0}^{2K-1} n^{m} (-1)^n h_{2K-1-n} \nonumber \\
        \label{eq:h_n_3rd_equation}
        = 0, \quad m < K .
    \end{gather}
\end{enumerate}
Solving Eqs.~(\ref{eq:h_n_1st_equation}),~(\ref{eq:h_n_2nd_equation}), and~(\ref{eq:h_n_3rd_equation}), yields two sets of solutions for the coefficients, $h_l$ and $h^{\prime}_l$. These solutions are related by $h_l = h^{\prime}_{2K-l-l}$, the associated scaling functions, $s(x)$ and $s^{\prime}(x)$, are mirror image of each other and have compact support over the interval $\left[0,2K-1\right]$. The values of the coefficient, $h_n$, for $K=3$ are presented in Table~\ref{tab:h_coefficients_K_3}.
\begin{table}[hbtp]
\begin{center}
\caption{\label{tab:h_coefficients_K_3}The values of the scaling coefficients, $h_n$, for $K=3$.} 
\setlength{\tabcolsep}{1.5pc}
\vspace{1mm}
\begin{tabular}{c | c }
\specialrule{.15em}{.0em}{.15em}
\hline
$h_n$ & $K=3$ \\
\hline
$h_0$ & $\dfrac{1+\sqrt{10}+\sqrt{5+2\sqrt{10}}}{16\sqrt{2}}$ \\
$h_1$ & $\dfrac{5+\sqrt{10}+3\sqrt{5+2\sqrt{10}}}{16\sqrt{2}}$ \\
$h_2$ & $\dfrac{10-2\sqrt{10}+2\sqrt{5+2\sqrt{10}}}{16\sqrt{2}}$ \\
$h_3$ & $\dfrac{10-2\sqrt{10}-2\sqrt{5+2\sqrt{10}}}{16\sqrt{2}}$ \\
$h_4$ & $\dfrac{5+\sqrt{10}-3\sqrt{5+2\sqrt{10}}}{16\sqrt{2}}$ \\
$h_5$ & $\dfrac{1+\sqrt{10}-\sqrt{5+2\sqrt{10}}}{16\sqrt{2}}$\\
\hline
\specialrule{.15em}{.15em}{.0em}
\end{tabular}
\end{center}
\end{table}
Given the values of $h_n$, Eq.~(\ref{eq:scaling_equation}) can be solved recursively to compute the values of the mother scaling function, $s(x)$, at each point $x$. From the mother scaling function, $s(x)$, a scaling function of resolution $k$ at position $m$ can then be obtained by applying the translation operator $m$ times, followed by the scale transformation operator $k$ times:
\begin{eqnarray}
    s^k_m(x) := \hat{D}^{k}\hat{T}^m s(x) =2^{k/2}s(2^k x-m).
\end{eqnarray}
Using the unitary property of $\hat{D}$ and $\hat{T}$, Eq.~(\ref{eq:orthonormality_of_the_mother_scfaling_function}) implies the orthonormality of the functions $s^k_m(x)$:
\begin{eqnarray}
\label{eq:orthonormality_of_scaling_functions}
    \int s^k_m(x) s^k_n(x) dx =\delta_{mn}.
\end{eqnarray}
The arbitrary linear combinations of the scaling functions $s^k_n(x)$ forms a resolution $k$ subspace $\mathcal{H}^k$ of the space of square integrable functions, defined by:
\begin{gather}
    \mathcal{H}^k := \left\{f(x)|f(x)=\sum_{n=-\infty}^{\infty} c_n s^k_n(x),\,\,\sum_{n=-\infty}^{\infty}|c_n|^2<\infty\right\}.
\end{gather}
It follows from Eq.~(\ref{eq:scaling_equation}) that the scaling subspaces are nested; that is, lower-resolution subspaces are proper subsets of higher-resolution subspaces:
\begin{eqnarray}
    \dots \mathcal{H}^{k-1}\subset \mathcal{H}^k \subset \mathcal{H}^{k+1}\subset \dots\, \,.
\end{eqnarray}
Hence, the space of square-integrable functions can be constructed from arbitrary linear combinations of scaling functions as the resolution tends to infinity:
\begin{eqnarray}
\label{eq:l_square_R_using_scaling_functions}
    L^2(\mathbb{R})=\lim_{n\rightarrow \infty}\mathcal{H}^k.
\end{eqnarray}
Because of the nested structure of the scaling subspaces, we define the orthogonal complement of $\mathcal{H}^k$ within $\mathcal{H}^{k+1}$, denoted by $\mathcal{W}^k$,
\begin{eqnarray}
    \mathcal{H}^{k+1}=\mathcal{H}^k\oplus\mathcal{W}^k.
\end{eqnarray}
The subspaces, $\mathcal{W}^k$, are called the wavelet spaces. By successively adding the wavelet subspaces at all higher resolutions, the space of square-integrable functions $L^2(\mathbb{R})$ can be expressed as the direct sum of scaling and wavelet subspaces:
\begin{eqnarray}
\label{eq:l^2_using_scaling_and_wavelet_functions}
    L^2(\mathbb{R}) = \mathcal{H}^k\oplus\mathcal{W}^k\oplus \mathcal{W}^{k+1}\oplus \mathcal{W}^{k+1}\dots
\end{eqnarray}
Similar to the scaling functions, the wavelet functions are also constructed from a single function, $w(x)$, called the mother wavelet function, which is again constructed by taking the weighted linear combination of the scaling function, $s(x)$, scaled to one halved of its support:
\begin{eqnarray}
    w(x):=\sum_{l=0}^{2K-1}g_l \hat{D}\hat{T}^l s(x).
\end{eqnarray}
The coefficients $g_l$ are related to the coefficients $h_l$ of Eq.~(\ref{eq:scaling_equation}) by the relation, $g_l=(-1)^l h_{2K-1-l}$.

Analogous to the construction of the scaling function of resolution $k$ at position $m$, a wavelet function of resolution $k$ at position $m$ is obtained by applying the translation operator $\hat{T}$ and the scale transformation operator $\hat{D}$, $m$ and $k$ times, respectively.
\begin{eqnarray}
    w^k_m(x):=\hat{D}^k\hat{T}^m w(x).
\end{eqnarray}
From Eq.~(\ref{eq:orthonormality_of_scaling_functions}), it can be shown that the wavelet functions are orthonormal to each other and are orthogonal to the scaling functions,
\begin{eqnarray}
    \int w^k_m(x) w^l_n(x)dx &=& \delta_{mn}\delta_{kl},\\
    \int s^k_m(x) w^l_n(x)dx &=& 0.
\end{eqnarray}
Now, there are two possible choices of bases of the subspace of resolution $k$: one is using the scaling functions of resolution $k$, $\left(\left\{s^{k}(x)\right\}_{n=-\infty}^{\infty}\right)$ and another is the combination of scaling and wavelet functions of resolution $k-1$, $\left(\left\{s^{k-1}(x)\right\}_{n=-\infty}^{\infty}\cup \left\{w^{k-1}(x)\right\}_{n=-\infty}^{\infty}\right)$. This bases are related through an orthogonal transformation:
\begin{gather}
    s^{k-1}_n(x) = \sum_{m=2n}^{2n+2K-1}H_{nm}s^k_{m}(x),\\
    w^{k-1}_n(x) = \sum_{m=2n}^{2n+2K-1}G_{nm}s^k_{m}(x),\\
    s^k_n(x)=\sum_{m=2n}^{2n+2K-1}H^t_{nm}s^{k-1}_{m}(x)+\sum_{m=2n}^{2n+2K-1}G^t_{nm}w^{k-1}_{m}(x),
\end{gather}
where,
\begin{gather}
\begin{gathered}
    H_{nm} = h_{m-2n}, \quad G_{nm} = g_{m-2n},\\
    H^t_{nm}=h_{n-2m}, \quad G^t_{nm} = g_{n-2m}.
\end{gathered}
\end{gather}
Daubechies wavelet functions are fractal in nature because the mother scaling function, from which all basis functions are generated, is the solution of the refinement equation, Eq.~(\ref{eq:scaling_equation}). Despite their fractal nature, these functions possess a finite number of derivatives that increases with the parameter $K$. In this work, we use $K=3$, since the order-3 Daubechies functions have one continuous derivative. Consequently, Hamiltonians containing at most one spatial derivative can be represented exactly using $K=3$ scaling functions. Increasing $K$ yields smoother functions, at the expense of larger support and increased overlap between basis functions. 
\section{The Flow equation method in Daubechies wavelet basis}
\label{sec:The_Flow_equation_method_in_Daubechies_wavelet_basis}
In this section, we outline the key elements of the flow-equation method introduced by Wegner~\cite{https://doi.org/10.1002/andp.19945060203}, and also discussed in our previous work~\cite{PhysRevD.111.096024}. In this method, the Hamiltonian $\mathrm{H}(0)$ is evolved to a unitary equivalent form, $\mathrm{H}(\lambda)$ using the unitary operator, $U(\lambda)$, which is parametrized by a continuous parameter, $\lambda$:
\begin{eqnarray}
\label{eq:the_hamiltonian_at_lambda}
\mathrm{H}(\lambda)=U(\lambda)\mathrm{H}(0)U(\lambda)
\end{eqnarray}
By taking the derivative of the both side of Eq.~(\ref{eq:the_hamiltonian_at_lambda}), we arrive at the following differential equation satisfied by $\mathrm{H}(\lambda)$:
\begin{eqnarray}
\label{eq:flow_eqn_Hamiltonian}
\frac{d\mathrm{H}(\lambda)}{d\lambda}=\left[K(\lambda),\mathrm{H}(\lambda)\right],
\end{eqnarray}
where, $K(\lambda)$ is the anti-hermitian generator of the unitary transformation. A suitable choice of the generator transforms the Hamiltonian to the desired form.

According to Polyzou~\cite{PhysRevD.95.094501}, to obtain the block diagonalized Hamiltonian from the initial Hamiltonian, the following generator has to be chosen,
\begin{eqnarray}
\label{eq:the_generator_of_the_flow}
K(\lambda)=\left[G(\lambda),\textrm{H}(\lambda)\right],
\end{eqnarray}
where, $G(\lambda)$ is the part of the Hamiltonian where different scale coupling terms are not present. As $G(\lambda)$ and $\mathrm{H}(\lambda)$ are hermitian, the commutation of these two matrices, $K(\lambda)$, is anti-hermitian.

\begin{figure*}[hbpt]
    \centering
    \caption{\label{fig:schematic_diagram_h_lambda}The schematic diagram of decomposition of the Hamiltonian, $\mathrm{H}(\lambda)$, into the same scale scale coupling matrix, $\mathrm{H}_b(\lambda)$ and different scale coupling matrix, $\mathrm{H}_c(\lambda)$ }
    \includegraphics[scale=0.7]{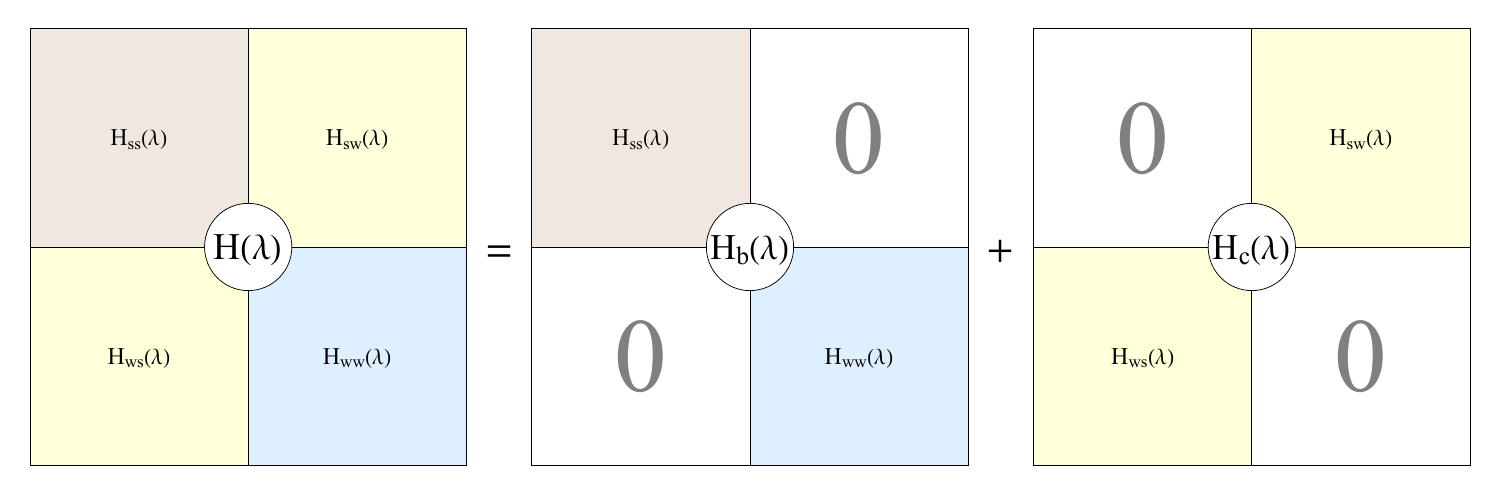}
\end{figure*}

For any arbitrary value of $\lambda$, the full Hamiltonian can be written as,
\begin{eqnarray}
\mathrm{H}(\lambda)=\mathrm{H}_b (\lambda)+\mathrm{H}_c (\lambda)
\end{eqnarray}
where, $\mathrm{H}_b(\lambda)$ is the same scale coupling term and $\mathrm{H}_c(\lambda)$ is the different scale coupling term. The schematic diagram of the above equation is depicted in Fig.~\ref{fig:schematic_diagram_h_lambda}. 

As elaborated in Ref.~\cite{PhysRevD.95.094501}, Eq.~(\ref{fig:schematic_diagram_h_lambda}) can be separated into two parts:
\begin{eqnarray}
\frac{d\mathrm{H}_{bmn}(\lambda)}{d\lambda}= \left(e_{cm}(\lambda)-e_{cn}(\lambda)\right)^2\mathrm{H}_{bmn}(\lambda),
\end{eqnarray}
and
\begin{eqnarray}
\frac{d\mathrm{H}_{cmn}(\lambda)}{d\lambda}= -\left(e_{bm}(\lambda)-e_{bn}(\lambda)\right)^2\mathrm{H}_{cmn}(\lambda),
\end{eqnarray}
where, $\mathrm{H}_{bmn}$ and $\mathrm{H}_{cmn}$ denote the matrix elements of the same scale and different scale coupling matrices respectively, whereas $e_{bn}$ and $e_{cn}$ are their corresponding eigenvalues. These equations can exactly be solved as,
\begin{eqnarray}
\mathrm{H}_{bmn}(\lambda) &=& e^{\int_{0}^{\lambda}\left(e_{cm}(\lambda')-e_{cn}(\lambda')\right)^2 d\lambda' }\mathrm{H}_{bmn}(\lambda),\\
\mathrm{H}_{cmn}(\lambda) &=& e^{-\int_{0}^{\lambda}\left(e_{bm}(\lambda')-e_{bn}(\lambda')\right)^2 d\lambda' }\mathrm{H}_{cmn}(\lambda).
\end{eqnarray}
One can observe from the above equations that the values of the matrix elements of $\mathrm{H}_{b}$ and $\mathrm{H}_c$ are exponentially increasing and decreasing, respectively, with the increasing value of the flow parameter $\lambda$. The evolution of the matrix elements ceases if there are degeneracies in the eigenvalues, approximate degeneracies, or if the eigenvalues intersect at the time of evolution. As the Hamiltonian matrix is infinite dimensional, to solve the equations numerically the Hamiltonian need to be truncated to a finite number of degrees of freedom. In order to do that, both the volume and resolution need to be truncated.

It is to be noted that, the choice of the generator of the flow is not unique. To obtain the scale-separated block diagonal form of the Hamiltonian, we choose the specific form of the generator given in Eq.~(\ref{eq:the_generator_of_the_flow}). Alternative choices of the generator, leading to different forms, are discussed in Ref.~\cite{PhysRevD.78.045011}.

\section{Wavelet discretized free scalar field theory}
\label{sec:wvaelet_discretized_free_scalar_field_theory}
In this section, we start with the review of the free scalar field theory conventions, then describe the formulation of the theory within the wavelet-based framework.

\subsection{Free scalar field theory: conventions}
The free scalar field theory Lagrangian in $1+1$ dimensions is given by,
\begin{eqnarray}
\mathcal{L}=\frac{1}{2} \int dx \left(\partial_{\mu}\phi\partial^{\mu}\phi - m^2 \phi^2 \right).
\end{eqnarray}
The corresponding Hamiltonian is given by,
\begin{eqnarray}
\label{eq:the_scalar_field_theory_hamiltonian}
\mathrm{H}=\frac{1}{2}\int dx \left(\pi^2 + (\nabla\phi)^2+ m^2 \phi^2 \right),
\end{eqnarray}
where, $\pi(x,t)$ is the canonical conjugate momentum of the field operator $\phi(x,t)$. The canonical conjugate pair of variables, $\phi(x,t)$ and $\pi(x,t)$, satisfies the usual equal time commutation relations:
\begin{eqnarray}
\begin{gathered}
\label{eq:the_cannonical_commutation_relation}
\left[\phi(x,t),\pi(y,t)\right]=i\delta(x-y),\\
 \left[\phi(x,t),\phi(y,t)\right]=\left[\pi(x,t),\pi(y,t)\right]=0.
\end{gathered}
\end{eqnarray}
The field operators in terms of the Fourier modes is given by,
\begin{gather}
\phi(x,t) = \int \frac{dp}{\sqrt{2(2\pi)2E}}\left(e^{i(Et-px)}a(p) + e^{-i(Et-px)}a^{\dagger}(p)\right),\\
\pi(x,t) = \int \frac{(iE)dp}{\sqrt{2(2\pi)2E}}\left(e^{i(Et-px)}a(p) - e^{-i(Et-px)}a^{\dagger}(p)\right),
\end{gather}
The Hamiltonian in terms of the Fourier modes is given by,
\begin{eqnarray}
\label{eq:Hamiltonian_momentum_creation_operator}
\mathrm{H}=\int dp E_p\left( a^{\dagger}(p)a(p) + \frac{1}{2}\delta(0)\right).
\end{eqnarray}
The second term is an infinite $c$-number. As we will see in the next section, a similar divergence term also appears in the case of wavelet-discretized fields, which can be removed by normal-ordering the Hamiltonian.
\subsection{Free scalar field theory: wavelet representation}
The field operators, $\left\{\phi(x,t),\pi(x,t)\right\}$, can be expanded in terms of the discrete scaling-wavelet fields operators, $\left\{\phi^{s,k}_n(t),\pi^{s,k}_n(t)\right\}$, as follows,
\begin{eqnarray}
\begin{gathered}
\label{eq:field_operators_in_terms_of_scaling_functions_and_wavelets}
\phi(x,t) = \sum_n \phi^{s,k}_n (t)s^k_n(x)+\sum_{l\geq k;n} \phi^{w,l}_{n}(t) w^l_n(x),\\
\pi(x,t) = \sum_n \pi^{s,k}_n (t)s^k_n(x)+\sum_{l\geq k;n} \pi^{w,l}_{n}(t) w^l_n(x).
\end{gathered}
\end{eqnarray}
where, scaling and wavelet field operators are defined as,
\begin{eqnarray}
\begin{gathered}
\phi^{s,k}_n(t) = \int dx \phi(x,t)s^k_n(x), \\
\phi^{w,l}_n(t) = \int dx \phi(x,t)w^l_n(x)\quad \left(l\geq k\right),\\
\pi^{s,k}_n(t) = \int dx \pi(x,t)s^k_n(x), \\
\pi^{w,l}_n(t) = \int dx \pi(x,t)w^l_n(x)\quad \left(l\geq k\right).
\end{gathered}
\end{eqnarray}
These new filed operators are smeared over their associated basis functions. From the commutation relations given in Eq.~(\ref{eq:the_cannonical_commutation_relation}), we derive the following commutation relations among the scaling and wavelet field variables:
\begin{eqnarray}
\begin{gathered}
\left[\phi^{s,k}_m(t),\phi^{s,k}_n(t)\right]=0,\quad \left[\phi^{s,k}_m(t),\phi^{w,l}_n(t)\right]=0,\\
\left[\phi^{s,k}_m(t),\pi^{s,k}_n(t)\right]=i\delta_{mn},\quad \left[\phi^{s,k}_m(t),\pi^{w,l}_n(t)\right]=0,\\
\left[\phi^{w,l}_m(t),\phi^{w,l}_n(t)\right]=0,\quad \left[\phi^{w,l}_m(t),\pi^{s,k}_n(t)\right]=0, \\
\left[\phi^{w,l}_m(t),\pi^{w,l'}_n(t)\right]=i\delta_{mn}\delta_{ll'},\quad \left[\pi^{s,k}_m(t),\pi^{s,k}_n(t)\right]=0,\\
\left[\pi^{s,k}_m(t),\pi^{w,l}_n(t)\right]=0,\quad \left[\pi^{w,l}_m(t),\pi^{s,k}_n(t)\right]=0.
\end{gathered} 
\end{eqnarray}
Substituting the expressions of the field operators given in Eq.~(\ref{eq:field_operators_in_terms_of_scaling_functions_and_wavelets}) into the Hamiltonian given in Eq.~(\ref{eq:the_scalar_field_theory_hamiltonian}), we arrive at the following form of the Hamiltonian,
\begin{eqnarray}
\label{eq:wavelet_discretized_free_scalar_field_theory_hamiltonian}
\mathrm{H}=\mathrm{H}_{ss}+\mathrm{H}_{ww}+\mathrm{H}_{sw}.
\end{eqnarray}
Here the first term,
\begin{eqnarray}
\begin{gathered}
\mathrm{H}_{ss}=\frac{1}{2}\left(\sum_n \pi^{s,k}_n(t)\pi^{s,k}_n(t) + \right.\\
\left. \sum_{m,n} \phi^{s,k}_m (t) D^{k}_{ss,mn}  \phi^{s,k}_m (t) + m^2 \sum_n \phi^{s,k}_n(t) \phi^{s,k}_n(t)\right),
\end{gathered}
\end{eqnarray}
denotes the part of the Hamiltonian consisting only of scaling field operators. The second term,
\begin{eqnarray}
\begin{gathered}
\mathrm{H}_{ww}=\frac{1}{2}\left(\sum_n \pi^{w,l}_n(t)\pi^{w,l}_n(t) + \right.\\
\left. \sum_{m,n} \phi^{w,l}_m (t) D^{lj}_{ww,mn}  \phi^{w,j}_m (t) + m^2 \sum_n \phi^{w,l}_n(t) \phi^{w,l}_n(t)\right),
\end{gathered}
\end{eqnarray}
denotes the part of the Hamiltonian consisting only of wavelet field operators, while the third term,
\begin{eqnarray}
\mathrm{H}_{sw} =\frac{1}{2}\sum_{m,n,l}\phi^{s,k}_m(t)D^{kl}_{sw,mn}\phi^{w,l}_n(t),
\end{eqnarray}
is the part of the Hamiltonian, where the interaction between scaling and wavelet field operators are present. The coefficients \( \mathcal{D}^{k}_{ss,mn} \), \( \mathcal{D}^{lj}_{ww,mn} \), and \( \mathcal{D}^{kl}_{sw,mn} \) represent the overlap integrals of the derivatives of the scaling functions at the same scale, the wavelet functions at the same and different scales, and the scaling and wavelet functions at different scales, respectively:
\begin{eqnarray}
\label{eq:overlap_integral_derivative_scaling_scaling}
\mathcal{D}^{k}_{ss,mn} &=& \int dx \, \frac{ds^k_m(x)}{dx} \frac{ds^k_n(x)}{dx}, \\
\label{eq:overlap_integral_derivative_wavelet_wavelet}
\mathcal{D}^{lj}_{ww,mn} &=& \int dx \, \frac{dw^l_m(x)}{dx} \frac{dw^j_n(x)}{dx}, \\
\mathcal{D}^{kl}_{sw,mn} &=& \int dx \, \frac{ds^k_m(x)}{dx} \frac{dw^l_n(x)}{dx} \label{eq:overlap_integral_derivative_scaling_wavelet}.
\end{eqnarray}
Note, the coefficient \( \mathcal{D}^{k}_{ss,mn} \) denotes the local and nearest-neighboring coupling of scaling modes at the same scale, \( \mathcal{D}^{lj}_{ww,mn} \) denotes the coupling of wavelet modes across the same and different scales, and \( \mathcal{D}^{kl}_{sw,mn} \) denotes the coupling between scaling and wavelet modes at different scales. Due to the compact support of the scaling and wavelet functions, the overlap integrals vanish when the difference in the translation indices of the associated functions exceeds a certain range. For example, in the case of order $3$ ($K=3$) Daubechies wavelet functions, the overlap integrals are non-zero only for $|m-n|<=4$. This property of the overlap integrals is similar to the finite difference approximation on a lattice, where the derivatives are replaced by differences of function values evaluated at discrete points separated by a small spacing.

Another important feature of these integrals is that they can be evaluated {\it {analytically}} using the properties of the scaling and wavelet functions. The details of the calculation are presented in Appendix~\ref{appen:overlap_integrals}. For a more comprehensive discussion, one may refer to Ref.~\cite{PhysRevD.87.116011,kessler2003waveletnotes,basak2025multiresolutionanalysisquantumtheories}.

\section{Effective Hamiltonian from Flow Equations}
\label{sec:Effective_Hamiltonian_from_Flow_Equations}
In this section, we apply the flow equation method to decouple the low- and high-scale degrees of freedom and derive the effective Hamiltonian, as detailed in Ref.~\cite{PhysRevD.111.096024}. The effective Hamiltonian is subsequently used for the second quantization of the theory. We begin by rewriting the wavelet discretized Hamiltonian of the free scalar field theory, Eq.~(\ref{eq:wavelet_discretized_free_scalar_field_theory_hamiltonian}), in matrix form: 
\begin{eqnarray}
\label{eq:Hamiltonian_matrix_form_scaling_wavelet}
\mathrm{H}= \frac{1}{2}\left[\Pi^{T}\Pi + \Phi^{T}\Omega^2 \Phi \right], 
\end{eqnarray}
where, $\Phi$ and $\Pi$ are the column matrices containing the field and the conjugate variables respectively,
\begin{eqnarray}
\Phi^{T} = \left(\Phi^{s,k}\quad \Phi^{w,l} \right)^T,\quad \Pi^{T} = \left(\Pi^{s,k}\quad \Pi^{w,l} \right)^T,
\end{eqnarray}
with,
\begin{eqnarray}
    \label{eq:phi_column_matrix_scaling_sector}
    \left(\Phi^{s,k}\right)^T &=& \left(\phi^{s,k}_{-\infty}\dots\phi^{s,k}_m\dots \phi^{s,k}_{\infty}\right)^T,\\
    \label{eq:phi_column_matrix_wavelet_sector}
    \left(\Pi^{s,k}\right)^T &=& \left(\pi^{s,k}_{-\infty}\dots\pi^{s,k}_m\dots \pi^{s,k}_{\infty}\right)^T,\\
    \left(\Phi^{w,l}\right)^T &=& \left(\phi^{w,l}_{-\infty}\dots\phi^{w,l}_m\dots \phi^{w,l}_{\infty}\right)^T,\\
    \left(\Pi^{w,l}\right)^T &=& \left(\pi^{w,l}_{-\infty}\dots\pi^{w,l}_m\dots \pi^{w,l}_{\infty}\right)^T,
\end{eqnarray}
where, $l\geq k$, and $k$ denotes the coarsest scale. $\Omega^2$ of Eq.~(\ref{eq:Hamiltonian_matrix_form_scaling_wavelet}) is the coupling matrix, containing the different mode coupling terms:
\begin{eqnarray}
\label{eq:full_omega_square_matrix}
\Omega^2 :=
\begin{pmatrix}
\mu^2 \delta_{mn} + \mathcal{D}^{k}_{ss,mn} & \mathcal{D}^{kq}_{sw,mn} \\
\mathcal{D}^{qk}_{ws,mn} & \mu^2 \delta_{mn} + \mathcal{D}^{lq}_{ww,mn}
\end{pmatrix}.
\end{eqnarray}
with, $\infty \leq m,n\leq \infty$.

As discussed in Ref.~\cite{PhysRevD.111.096024}, the flow equation for the Hamiltonian reduces to that for the $\Omega^2$ matrix:
\begin{eqnarray}
\label{eq:omega^2_flow_equation}
\frac{d\Omega^2(\lambda)}{d\lambda}=\left[K(\lambda),\Omega^2(\lambda)\right].
\end{eqnarray}
To obtain the decoupled block-diagonal form, we chose the generator as given in Eq.~(\ref{eq:the_generator_of_the_flow}),
\begin{eqnarray}
K(\lambda)=\left[G(\lambda),\Omega^2(\lambda)\right],
\end{eqnarray}
where, $G(\lambda)$ is the matrix consists of the diagonal blocks of $\Omega^2(\lambda)$, in which the different scale-coupling terms are absent.

\begin{figure}[hbpt]
    \centering
    \caption{\label{fig:schematic_diagram_omegasquarelambda_res_2}The schematic diagram of the $\Omega^2(\lambda)$ matrix for resolution $2$.}
    \includegraphics[scale=0.55]{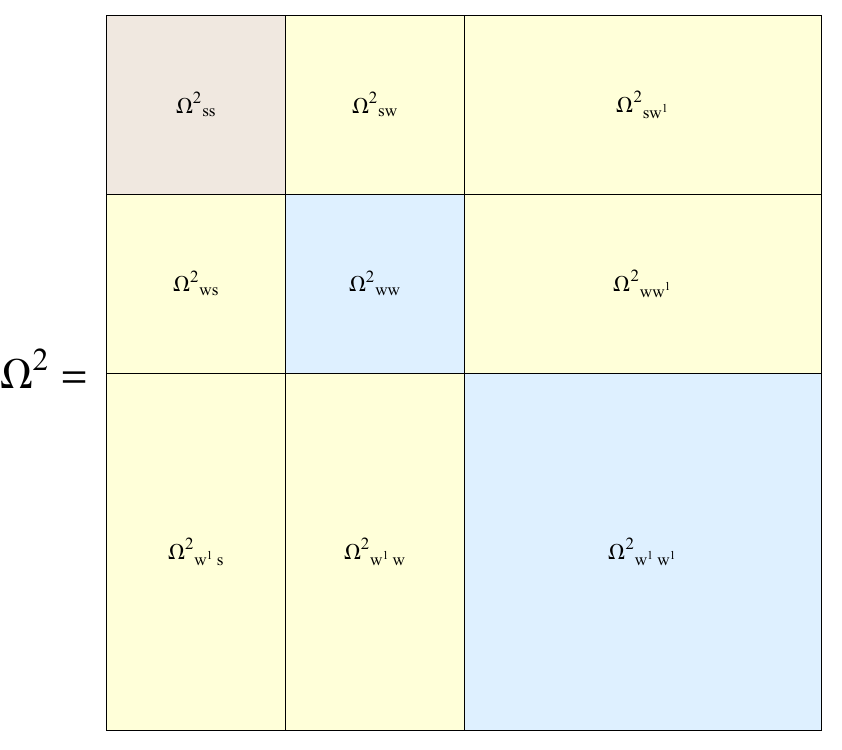}
\end{figure}

Now, the $\Omega^2$ matrix is constructed within the subspaces truncated up to resolution $4$, formed by the direct sum of resolution $0$ scaling and wavelet functions at resolutions $0$, $1$, $2$, and $3$. The space is truncated to $0\leq x\leq 10$. We impose the periodic boundary conditions to construct the matrix $\Omega^2$. The schematic diagram of $\Omega^2$ matrix for resolution $2$ is presented in Fig.~\ref{fig:schematic_diagram_omegasquarelambda_res_2}.

\begin{figure*}[hbtp]
    \centering
    \caption{\label{fig:hamiltonian_increasing_lambda_res_2}The matrix plot of the effective Hamiltonian of resolution $2$ with the increasing values of flowing parameter $\lambda$}
    \includegraphics[scale=0.35]{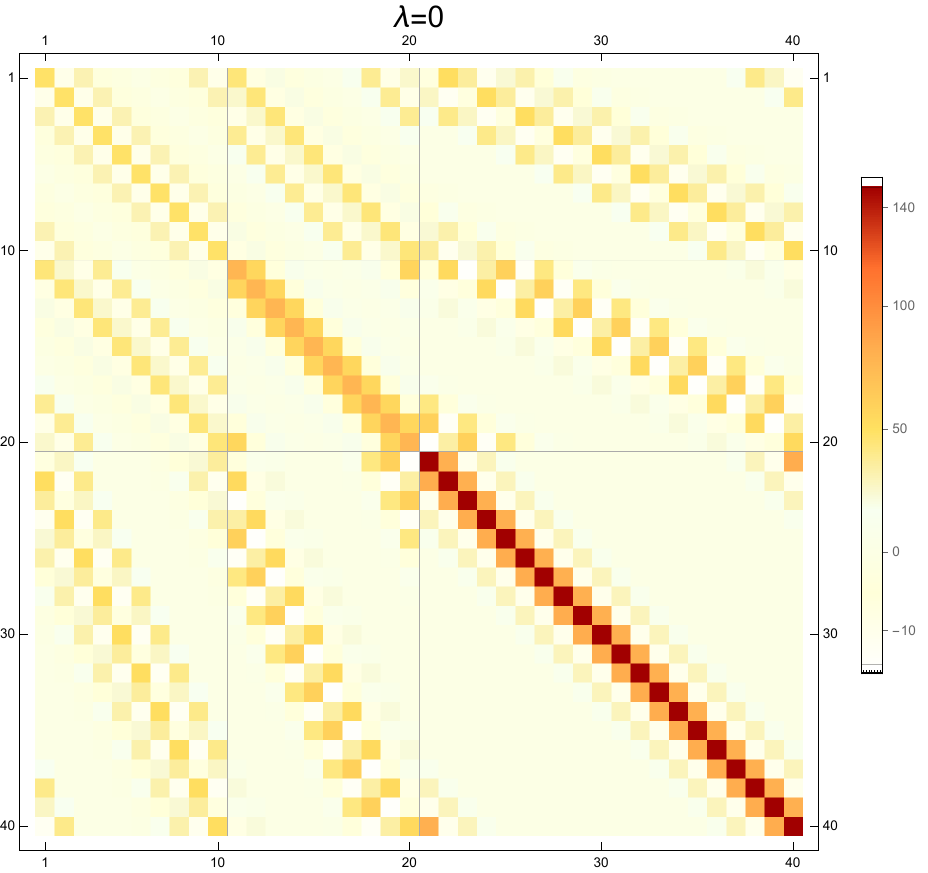}
    \includegraphics[scale=0.35]{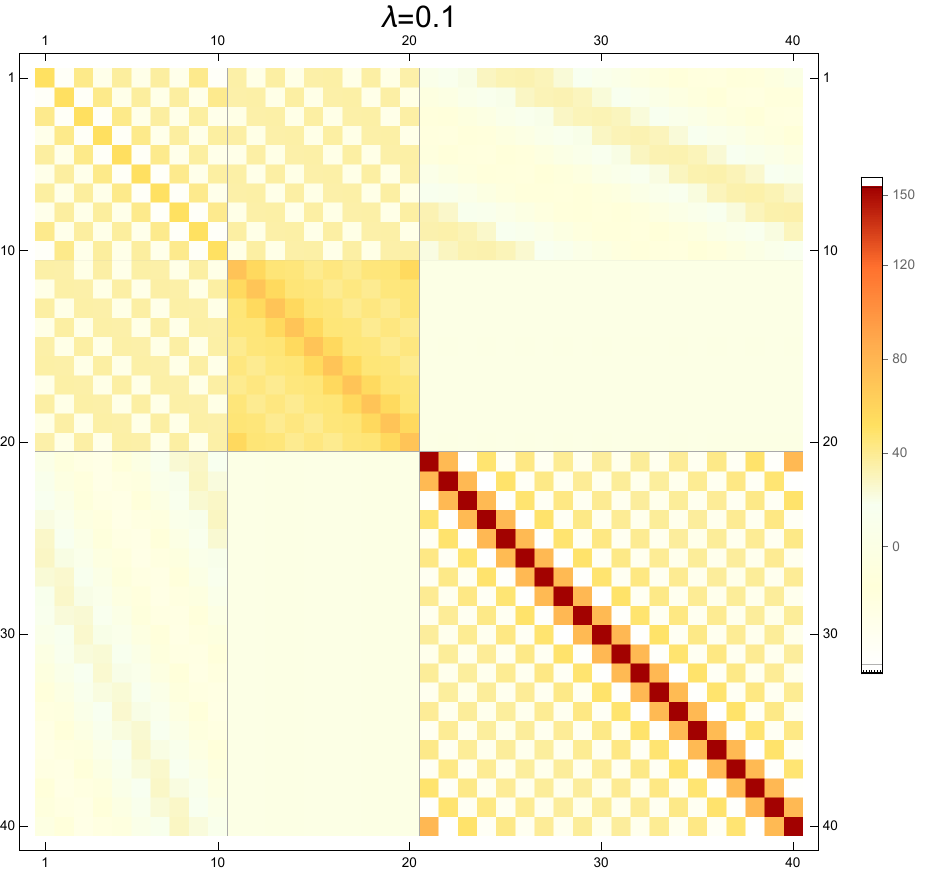}
    \includegraphics[scale=0.35]{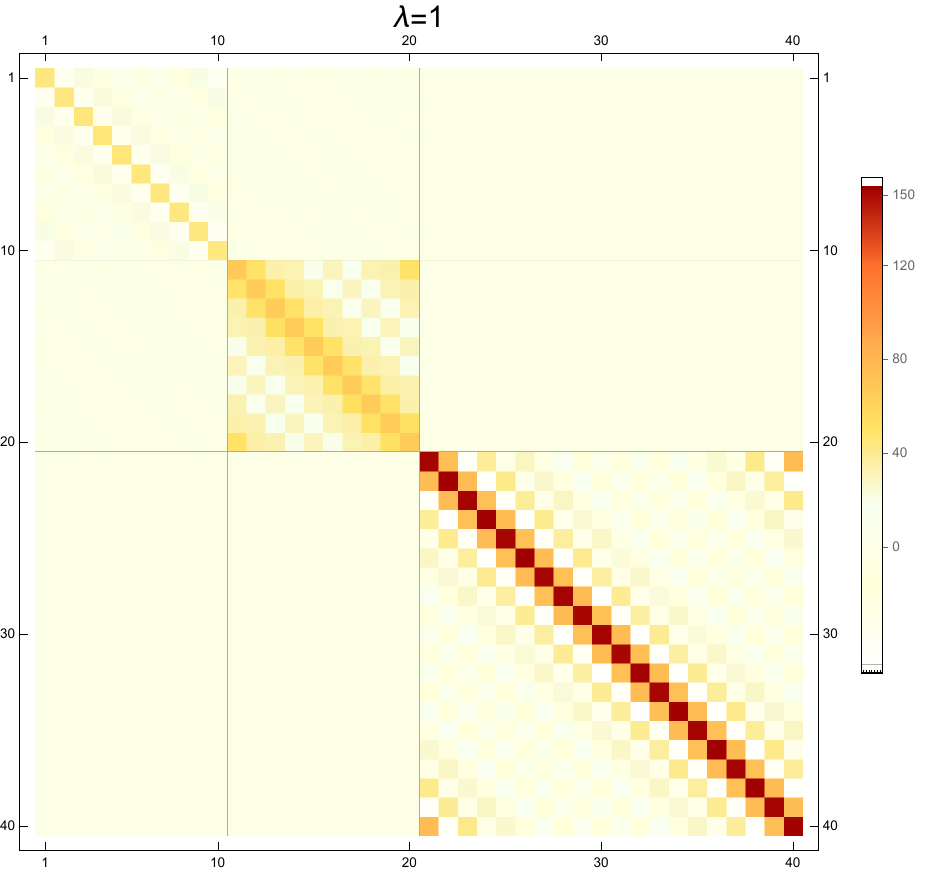}
    \includegraphics[scale=0.35]{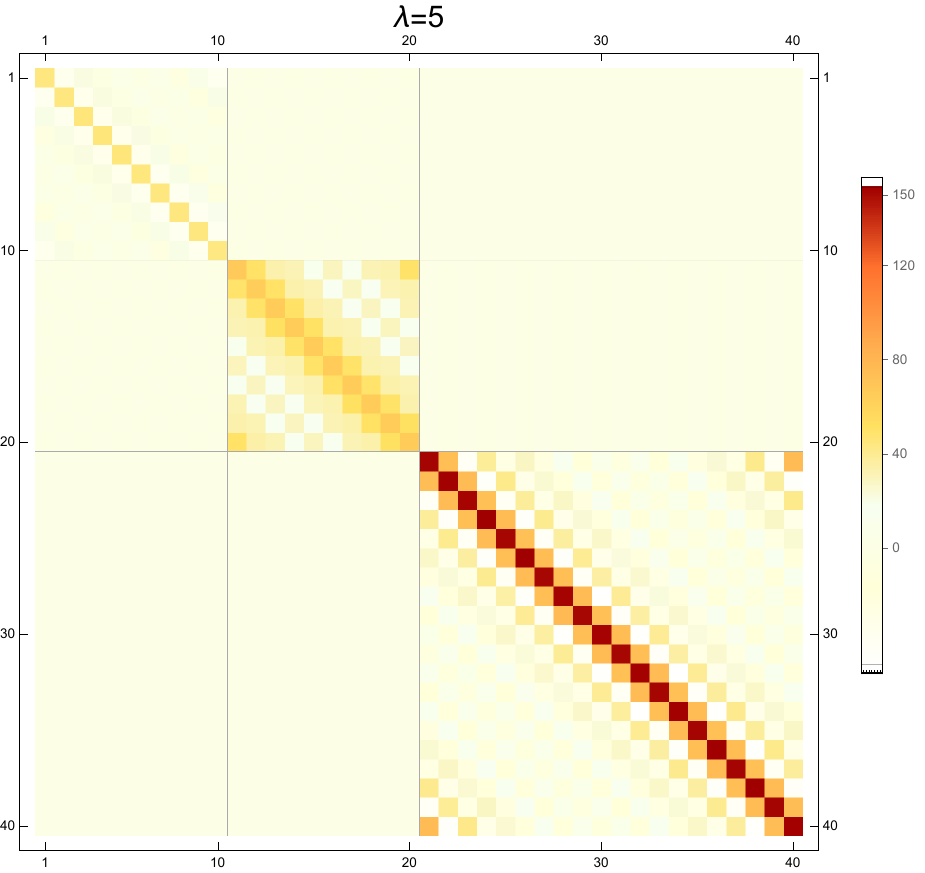}
    \includegraphics[scale=0.35]{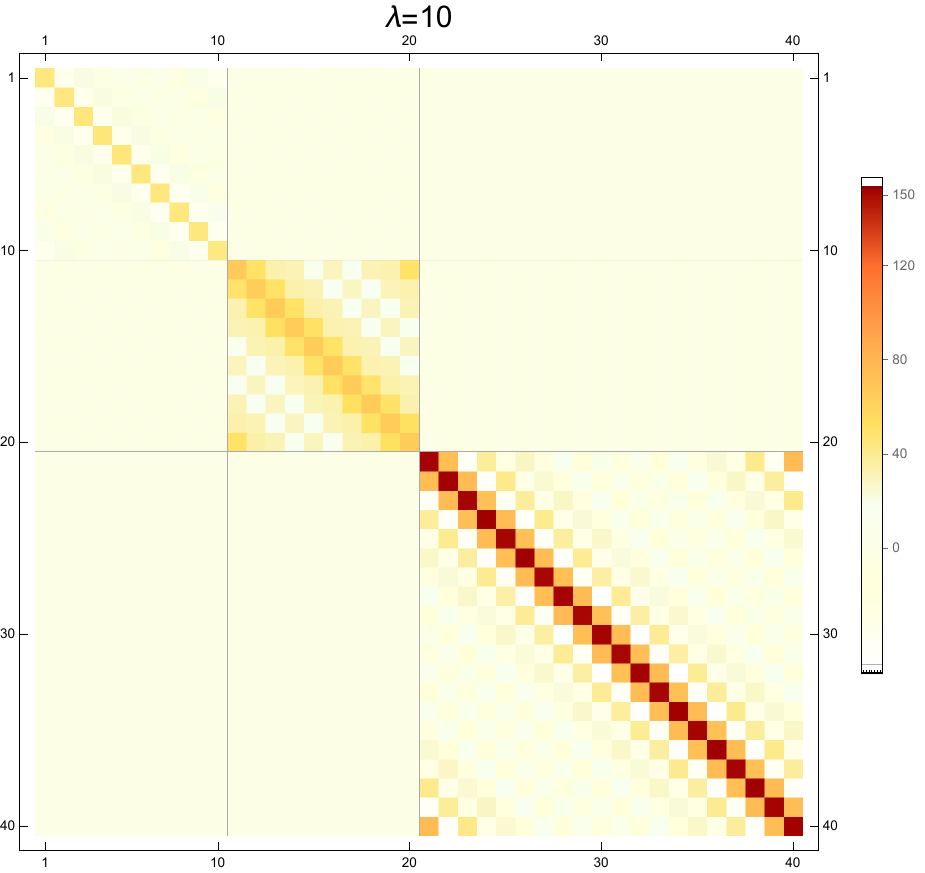}
    \includegraphics[scale=0.35]{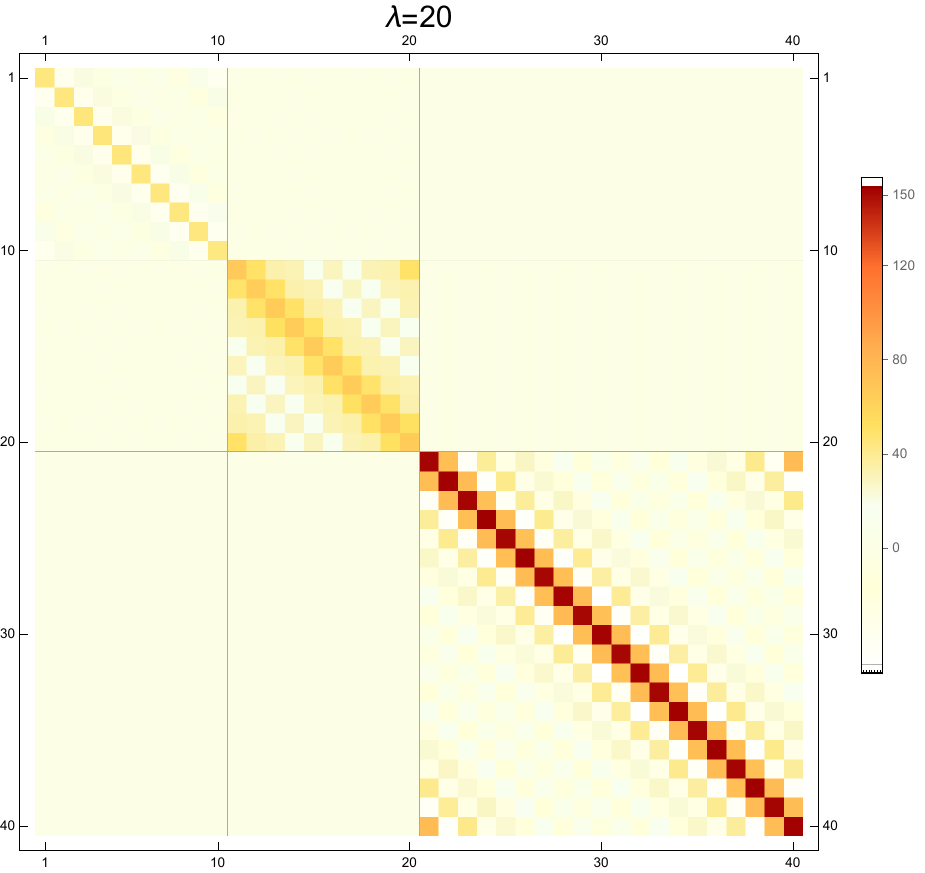}
\end{figure*}

For this choice of $\Omega^2(\lambda)$, we solve Eq.~(\ref{eq:omega^2_flow_equation}) for the increasing values of $\lambda$. To illustrate the evolution of the matrix with increasing values of $\lambda$, the resulting matrix plots for resolution $2$ are depicted in Fig.~\ref{fig:hamiltonian_increasing_lambda_res_2}. The other matrix plots of the resulting matrix for increasing values of $\lambda$ with resolution $1$, $3$, and $4$ are presented in Appendix~\ref{appen:the_matrix_plots}. In Table~\ref{tab:eigenvalues_scalar_field_theory_omega_square_increasing_lambda}, we present the eigenvalues of $\Omega^2(\lambda)$ for increasing values of $\lambda$ corresponding to $0$, $0.1$, $1$, $5$, $10$ and $20$.
\begin{table*}[hbtp]
\begin{center}
\caption{Comparison among the $\omega^2_p$s, calculated analytically, obtained by diagonalizing the full truncated $\Omega^2(\lambda)$ matrix, and obtained by diagonalizing the scaling-scaling sector of the $\Omega^2$ matrix for the increasing values of $\lambda$, for resolution $2$} 
\label{tab:eigenvalues_scalar_field_theory_omega_square_increasing_lambda}
\setlength{\tabcolsep}{.8pc}
\vspace{1mm}
\begin{tabular}{c | c | c | c | c | c | c | c}
\specialrule{.15em}{.0em}{.15em}
\hline
\multicolumn{8}{c}{Comparison of the eigenvalues of $\Omega_p^2$ for resolution $2$ ($k=2$)}\\
\hline
Analytical & Full truncated & $\lambda=0$ & $\lambda=0.1$ & $\lambda=1$ & $\lambda=5$ & $\lambda=10$ & $\lambda=20$\\
\hline
$1.000000$ & $1.000000$ & $1.000000$  & $1.000000$ & $1.000000$ & $1.000000$ & $1.000000$ & $1.000000$\\
$1.394784$ & $1.394819$ & $1.402767$  & $1.394819$ & $1.394819$ & $1.394819$ & $1.394819$ & $1.394819$\\
$2.579137$ & $2.581318$ & $2.936329$  & $2.581318$ & $2.581318$ & $2.581318$ & $2.581318$ & $2.581318$\\
$4.553058$ & $4.577000$ & $6.757352$  & $4.577000$ & $4.577000$ & $4.577000$ & $4.577000$ & $4.577000$\\
$7.316547$ & $7.444279$ & $12.233314$ & $7.444279$ & $7.444279$ & $7.444279$ & $7.444279$ & $7.444279$\\
$10.869604$ & $11.325415$ & $15.019048$ & $11.325415$ & $11.325415$ & $11.325415$ & $11.325415$ & $11.325415$\\
\hline
\specialrule{.15em}{.15em}{.0em}
\end{tabular}
\end{center}
\end{table*}
One can observe from the results showed in Table.~\ref{tab:eigenvalues_scalar_field_theory_omega_square_increasing_lambda}, that the eigenvalues of the scaling-scaling sector of $\Omega^2(\lambda)$ matrix are converging towards the eigenvalues of the full-truncated matrix. However, noticeable deviations from the analytically calculated eigenvalues remain due to the truncation to a significantly lower resolution. As demonstrated in Ref.~\cite{PhysRevD.111.096024}, achieving closer agreement with the analytical values requires increasing the resolution to higher values, which is illustrated down below.

Table.~\ref{tab:eigenvalues_of_the_scaling_scaling_sector_of_lambda_20}, exhibits the eigenvalues of the scaling-scaling sector of the matrix $\Omega^2$ with the increasing values of the resolution $k$ for $\lambda=20$.
\begin{table*}[hbtp]
\begin{center}
\caption{\label{tab:eigenvalues_of_the_scaling_scaling_sector_of_lambda_20}Comparisons of the eigenvalues of the scaling-scaling sector of $\Omega^2(\lambda)$ with the increasing values of the resolution, $k$, for $\lambda=20$.} 
\setlength{\tabcolsep}{.8pc}
\vspace{1mm}
\begin{tabular}{c | c | c | c | c | c}
\specialrule{.15em}{.0em}{.15em}
\hline
\multicolumn{6}{c}{Comparison of the eigenvalues of the scaling-scaling sector of $\Omega_p^2$ for $\lambda=20$}\\
\hline
Analytical & $k=0$ & $k=1$ & $k=2$ & $k=3$ & $k=4$ \\
\hline
$1.000000$  & $1.000000$  & $1.000000$  & $1.000000$  & $1.000000$ & $1.000000$\\
$1.394784$  & $1.402767$  & $1.395329$  & $1.394819$  & $1.394786$ & $1.394784$\\
$2.579137$  & $2.936329$  & $2.611069$  & $2.581317$  & $2.579276$ & $2.579145$\\
$4.553058$  & $6.757352$  & $4.866601$  & $4.577000$  & $4.554631$ & $4.553157$\\
$7.316547$  & $12.233314$ & $8.745316$  & $7.444279$  & $7.325271$ & $7.317105$\\
$10.869604$ & $15.019048$ & $15.019048$ & $11.325415$ & $10.902335$ & $10.871722$\\
\hline
\specialrule{.15em}{.15em}{.0em}
\end{tabular}
\end{center}
\end{table*}
Table.~\ref{tab:eigenvalues_of_the_scaling_scaling_sector_of_lambda_20} shows that, with the increasing values of the resolution the eigenvalues of the scaling-scaling sector of the Hamiltonian is approaching towards the analytically calculated eigenvalues for $\lambda=20$. 

The above discussion clearly indicates that, in formulating the second quantization of scalar field theory within the Daubechies wavelet framework, it is advantageous to use only the scaling–scaling sector of the $\Omega^2$ matrix—rather than the full truncated matrix—to compute the low-lying eigenvalues. This approach substantially reduces the number of degrees of freedom and, consequently, the computational cost. We elaborate on this in the next section.

\section{Second Quantization using Daubechies wavelet basis: Scalar field theory}
\label{sec:second_quantization}
In this section, to obtain the low-lying eigenvalues, we introduce the wavelet-based second quantization of the scalar field theory by defining effective scaling creation and annihilation operators. We then construct the corresponding Fock basis and compute the Hamiltonian matrix elements within this effective scaling Fock basis. Finally, we present numerical results for the low-lying energy spectrum with the increasing resolution.

\subsection{Scaling creation and annihilation operator}
We begin by rewriting the scaling-scaling sector of the effective Hamiltonian for the free scalar field theory as follows:
\begin{eqnarray}
\label{eq:Hamiltonian_matrixform_only_scaling}
\mathrm{H}_{ss}= \frac{1}{2}\left[\left(\Pi^{s,k}\right)^T\Pi^{s,k} + \left(\Phi^{s,k}\right)^{T}\Omega^2_{ss} \Phi^{s,k} \right], 
\end{eqnarray}
where, $\Phi^{s,k}$ and $\Pi^{s,k}$ column vectors containing the scaling field coefficients and their conjugate momenta, respectively, as defined in Eqs.~(\ref{eq:phi_column_matrix_scaling_sector}) and~(\ref{eq:phi_column_matrix_wavelet_sector}). The scaling mode coupling matrix $\Omega^2_{ss}$ is given by,
\begin{gather}
\Omega^2_{ss}:=\nonumber\\
\label{eq:the_omega_square_scaling_representation}
\begin{pmatrix}
\mu^2+\mathcal{D}^{k}_{ss,-\infty -\infty} & \cdots & 
\mathcal{D}^{k}_{ss,-\infty m} & \cdots & \mathcal{D}^{k}_{ss,-\infty \infty} \\
\vdots & \ddots & & & \vdots \\
\mathcal{D}^k_{ss,m-\infty} & \cdots & \mu^2+\mathcal{D}^k_{ss,mm} & \cdots & \mathcal{D}^k_{ss,m\infty}\\
\vdots & & & \ddots & \vdots \\
\mathcal{D}^k_{ss,\infty -\infty} & \cdots & \mathcal{D}^k_{ss,\infty m} & \cdots & \mu^2+ \mathcal{D}^k_{ss,\infty\infty}
\end{pmatrix}.
\end{gather}
Eq.~(\ref{eq:Hamiltonian_matrixform_only_scaling}) is analogous to the Hamiltonian of a single Harmonic oscillator with a quadratic potential. We therefore introduce the corresponding creation and annihilation operator column matrices as,
\begin{eqnarray}
a^{s,k} &=& \frac{1}{\sqrt{2}}\left(A_{ss}\Phi^{s,k}+ i B_{ss}\Pi^{s,k} \right),\\
a^{s,k^{\dagger}} &=& \frac{1}{\sqrt{2}}\left(A_{ss}\Phi^{s,k}- i B_{ss}\Pi^{s,k} \right),
\end{eqnarray}
where, $A_{ss}=\sqrt{\Omega_{ss}}, B_{ss}=\frac{1}{\sqrt{\Omega_{ss}}}=\frac{1}{A_{ss}}$. $\Phi^{s,k}$, and $\Pi^{s,k}$ are the field operator column matrices given in the Eqs.~(\ref{eq:phi_column_matrix_scaling_sector}) and~(\ref{eq:phi_column_matrix_wavelet_sector}), respectively, and the matrix $\Omega_{ss}$ is the square root of the matrix $\Omega^2_{ss}$ given in Eq.~(\ref{eq:the_omega_square_scaling_representation}). Due to the positive definite nature of the $\Omega^2_{ss}$-matrix, the matrix $\Omega_{ss}$ is well-defined and thus the two matrices $A_{ss}$ and $B_{ss}$ are also well-defined. It is also possible to define the annihilation and the creation operator corresponding to the scaling and wavelet mode using the full $\Omega^2$ matrix given in Eq.~(\ref{eq:full_omega_square_matrix}) before applying the flow-equation transformation, which leads to an exponential increase in the dimensions of the Hamiltonian matrix with increasing resolution. However, as discussed in the previous section, the flow-equation method block-diagonalizes the $\Omega^2$ matrix, yielding effective scaling degrees of freedom corresponding to the $\Omega^2_{ss}$ sector of the block-diagonalized $\Omega^2$ matrix~(see the schematic diagram Fig.~\ref{fig:schematic_diagram_omegasquarelambda_res_2}) that incorporate the effects of higher-resolution wavelet modes. Consequently, the low-lying eigenvalues at higher resolutions can be obtained by working only with these effective scaling–scaling sector, $\Omega_{ss}$, of the matrix $\Omega$ obtained from square root of the block-diagonalized $\Omega^2$.  The operator column matrices $a^{s,k}$ and $a^{s,k^{\dagger}}$ contains a set of annihilation and creation operators corresponding to the effective scaling mode given by,
\begin{eqnarray}
\label{eq:scaling_creation}
a^{s,k}_m &:=& \frac{1}{\sqrt{2}}\sum_{n}\left[A_{ss,mn}\phi^{s,k}_n +iB_{ss,mn}\pi^{s,k}_n \right],\\
a^{s,k^\dagger}_m &:=& \frac{1}{\sqrt{2}}\sum_{n}\left[A_{ss,mn}\phi^{s,k}_n - iB_{ss,mn}\pi^{s,k}_n\right].
\end{eqnarray}
These annihilation and creation operators respectively annihilate and create an excitation of resolution $k$ located at position $m$ associated with the scaling mode. As demonstrated in Sec. \ref{sec:wvaelet_discretized_free_scalar_field_theory}, the overlap integrals, $\mathcal{D}^{k}_{ss,mn}$, contained within the scaling-scaling sector, $\Omega^2_{ss}$, of the matrix $\Omega^2$ are non-zero only for a finite range of the difference in the translation index of the two scaling functions involved. This leads to the band diagonal structure of the block $\Omega^2_{ss}$ of the matrix $\Omega^2$. The band-diagonal structure of $\Omega^2_{ss}$ remain preserved for sufficiently large values of the flow parameter $\lambda$~($\lambda=20$) under the similarity transformation, although the bandwidth is modified by the flow. The square root of $\Omega^2_{ss}$, i.e. $\Omega_{ss}$ is not strictly band-diagonal, the values of the matrix elements decays rapidly as one moves away from the diagonal elements. Hence, the matrices $A_{ss}$ and $B_{ss}$, defined as the square root and the inverse square root of the matrix $\Omega_{ss}$, respectively, exhibit the structure similar to that of the matrix $\Omega_{ss}$. Therefore, the excitation created by the creation operators is quasi-localized in the position space both before and after the flow-equation transformation with amplitudes that decay rapidly away from their center at $m$.

The inverse of these relations can be written in terms of the column matrices corresponding to the field and its conjugate momentum as: 
\begin{gather}
\label{eq:Phi_in_terms_of_a_a_dagger}
    \Phi^{s,k} = \frac{B_{ss}}{\sqrt{2}}\left(a^{s,k^{\dagger}} + a^{s,k}\right) = \frac{1}{\sqrt{2\Omega_{ss}}}\left( a^{s,k^{\dagger}} + a^{s,k}\right),\\
\label{eq:Pi_in_terms_of_a_a_dagger}
    \Pi^{s,k} = i\frac{A_{ss}}{\sqrt{2}}\left(a^{s,k^{\dagger}} - a^{s,k}\right) = i\sqrt{\frac{\Omega_{ss}}{2}}\left( a^{s,k^{\dagger}} - a^{s,k}\right),
\end{gather}
where $a^{s,k}$ and $a^{s,k^{\dagger}}$ are the creation and annihilation operator column matrices, respectively. Substituting $\Phi^{s,k}$ and $\Pi^{s,k}$ of Eq.~(\ref{eq:Hamiltonian_matrixform_only_scaling}) with the expressions of these operators given in Eqs.~(\ref{eq:Phi_in_terms_of_a_a_dagger}) and~(\ref{eq:Pi_in_terms_of_a_a_dagger}), we obtain the following form of the Hamiltonian,
\begin{eqnarray}
    \mathrm{H}_{ss}=(a^{s,k^{\dagger}})^T\Omega_{ss} a^{s,k} + \frac{1}{2}Tr[\Omega_{ss}].
\end{eqnarray}
The untruncated $\Omega_{ss}$ matrix is infinite  dimensional resulting the value of the second term of the right hand side of the above expression is infinity, which represents the zero-point energy of the system. This is analogous to the second term in the expression of the Hamiltonian given in Eq.~(\ref{eq:Hamiltonian_momentum_creation_operator}) within the present formalism. The normal-ordered Hamiltonian can hence be obtained by rescaling the zero point energy (to zero). The resulting Hamiltonian, in terms of scaling and wavelet creation and annihilation operator, can then be expressed as,
\begin{eqnarray}
    \textrm{H}_{ss} = \sum_{m,n}\Omega^{k}_{ss,mn}a^{s,k^{\dagger}}_m a^{s,k}_n .
\end{eqnarray}
As already discussed, the values of the matrix elements, $\Omega^k_{ss,mn}$, of $\Omega_{ss}$ decays rapidly as one moves away from the diagonal elements. This property of the matrix $\Omega_{ss}$ implies that the interaction strength between the scaling mode at position $m$ and another scaling mode is maximal when both are located at $m$, and it decays rapidly as the second scaling mode moves away from $m$. 

\subsection{The Fock space basis construction in wavelet basis}
We define the Fock space basis element using the scaling modes as,  
\begin{eqnarray}
    S=\ket{m1^{s,k}_{-\infty}\dots m1^{s,k}_{p}\dots m1^{s,k}_{\infty}}.
\end{eqnarray}
Here, $m1^{s,k}_p$ denotes the the occupation numbers of the scaling mode with the translation and resolution indices $p$ and $k$ respectively. The action of the creation and annihilation operators, given in Eq.~(\ref{eq:scaling_creation}), on these states is given by,
\begin{eqnarray}
\begin{gathered}
    a^{s,k\dagger}_p\ket{m1^{s,k}_{-\infty}\dots m1^{s,k}_{\infty}}\\
     = \sqrt{m1^{s,k}_p +1}\ket{m1^{s,k}_{-\infty}\dots (m1^{s,k}_{p}+1)\dots m1^{s,k}_{\infty}},
\end{gathered}\\
\begin{gathered}
    a^{s,k}_p\ket{m1^{s,k}_{-\infty}\dots m1^{s,k}_{\infty}} \\
    = \sqrt{m1^{s,k}_p}\ket{m1^{s,k}_{-\infty}\dots (m1^{s,k}_{p}-1)\dots m1^{s,k}_{\infty}}.
\end{gathered}
\end{eqnarray}
After defining the action of the creation and the annihilation operators on the basis states, we can now proceed to construct the Hamiltonian matrix, which is given below,
\begin{widetext}
\begin{eqnarray}
\mathrm{H}_{ss,m1m2}&&=\bra{m1^{s,k}_{-\infty},\ldots,m1^{s,k}_{\infty}}\mathrm{H}\ket{m2^{s,k}_{-\infty},\ldots, m2^{s,k}_{\infty}},\nonumber\\
&&
\label{eq:free_hamiltonian_matrix_element}
\begin{cases}
=\sum_{p,q}T^{k}_{ss,pq}\delta_{m1^{s,k}_{-\infty},m2^{s,k}_{-\infty}}\ldots \delta_{m1^{s,k}_{\infty},m2^{s,k}_{\infty}},\,\, \text{for}\,\, p= q\\
=\sum_{p,q}T^{k}_{ss,pq}\delta_{m1^{s,k}_{-\infty},m2^{s,k}_{-\infty}}\ldots \delta_{m1^{s,k}_p,m2^{s,k}_p+1}\ldots \delta_{m1^{s,k}_q,m2^{s,k}_q-1}\ldots \delta_{m1^{s,k}_{\infty},m2^{s,k}_{\infty}},\,\, \text{for}\,\, p\neq q ,
\end{cases}
\end{eqnarray}
where,
\begin{eqnarray}
T^{k}_{ss,pq}=
\begin{cases}
\Omega^{k}_{ss,pp}m2^{s,k}_p \,\, \text{for}\,\, p=q\\
\Omega^{k}_{ss,pq}\sqrt{(m2^{s,k}_p+1)m2^{s,k}_q}\,\,\text{for}\,\, p\neq q .
\end{cases}
\end{eqnarray}
\end{widetext}
\subsection{Truncation scheme and the energy spectrum}
To construct a finite-dimensional Hamiltonian matrix, we impose both ultraviolet and infrared cutoffs on the degrees of freedom of the system. The ultraviolet cutoff is implemented by limiting the resolution index, while the infrared cutoff is introduced by restricting the translation index. In addition, the occupation number of each mode is bounded by a maximum value, $N_{\text{max}}$. Together, these constraints correspond to working within a truncated subspace of the Hilbert space defined by the chosen resolution.

We begin the construction of the finite-size Fock space basis within the zero-resolution subspace by restricting the spatial domain to $0 \leq x \leq 10$, imposing a maximum occupation number $N_{\text{max}} = 10$, and applying periodic boundary conditions. This spatial truncation corresponds to limiting the translation index to $0 \leq m \leq 6$. Periodic boundary conditions are implemented by extending the $\Omega$ matrix through periodic repetition of its elements, such that the 5th element is identified with the 7th, the 4th with the 8th, the 3rd with the 9th, and the 2nd with the 10th. Consequently, each Fock basis element contains 10 modes, resulting in a total Fock space dimension of 184,756.

The Hamiltonian commutes with the momentum operator $\mathrm{P}$, i.e., $\left[\mathrm{H},\mathrm{P}\right]=0$. Consequently, the Hamiltonian can be block-diagonalized into sectors of definite momentum, and the low-lying eigenvalues are obtained from the zero-momentum sector. In Appendix~\ref{appen:zero_momentum} we illustrate the procedure of projection of the Fock basis states onto the zero momentum sector. The number of Fock basis states associated to the zero momentum sector is $18504$.

Furthermore, the Hamiltonian commutes with the spatial parity operator ($\mathbb{P}$), i.e, $\left[\mathrm{H},\mathbb{P}\right]=0$ , under which the Lagrangian is invariant for the space transformation, $x \rightarrow -x$. Hence, the zero momentum sector further splits into even parity ($\mathbb{P}=+1$) and the odd parity ($\mathbb{P}=-1$) sectors. The procedure of taking projection of the basis states onto the $\mathbb{P}=+1$ and $\mathbb{P}=-1$ sector is detailed in the Appendix~\ref{appen:parity_projection}. The number of basis states associated with these sectors thus becomes $9588$ and $8916$, respectively. In this work, we compute the eigenvalues of zero momentum ($\mathrm{P}=0$) and even parity ($\mathbb{P}=+1$) sector of the Hamiltonian.

Next, the ultraviolet cutoff is systematically increased by increasing the resolution. As discussed in Sec.~\ref{sec:Effective_Hamiltonian_from_Flow_Equations}, the effect of the higher resolution degrees of freedom can be encapsulated within the smaller set of degrees of freedom by block-diagonalizing the $\Omega^2$ matrix using the flow equations. 
To construct the effective higher-resolution Fock space basis, first we consider the leftmost block of the $\Omega^2$ matrix, denoted as $\Omega^2_{ss}$ in Fig.~\ref{fig:schematic_diagram_omegasquarelambda_res_2}. This block represents the effective scaling sector for $\lambda = 20$. The corresponding $\Omega_{ss}$ matrix is then derived from this effective $\Omega^2_{ss}$ and utilized to build the higher-resolution Fock space basis. Notably, the number of basis elements in the higher-resolution space remains unchanged from the lower-resolution case. The low-lying eigenvalues in the $\mathrm{P}=0$ and $\mathbb{P}=+1$ sector of the Hamiltonian, obtained with increasing resolution, are presented in Table.~\ref{tab:eigenvalues_of_the_scalar_field_theory_increasing_resolution}.

In Table.~\ref{tab:eigenvalues_of_the_scalar_field_theory_increasing_resolution}, we present results at different resolutions. It is evident that the eigenvalues converge toward their exact values (shown in the first column) as the resolution increases (other columns). For instance, the eigenvalue 2.362020, corresponding to a zero-momentum two-particle state with momenta $+2$ and $-2$, has an accuracy of $99.714\%$ at resolution $k=0$, which improves to agreement up to six decimal places by $k=4$. Similarly, for the two-particle state with momenta $+4$ and $-4$, the exact value is $3.211938$. At resolution $k=0$, the accuracy is $93.72\%$, which improves to $99.99\%$ as the  resolution increases to $k=4$.

\begin{table*}[!t]
\begin{center}
\caption{\label{tab:eigenvalues_of_the_scalar_field_theory_increasing_resolution}Comparisons of the eigenvalues of the zero momentum ($\hat{\mathrm{P}}=0$) and the even parity ($\hat{\mathbb{P}}=+1$) sector of the Hamiltonian with the increasing resolutions.}
\setlength{\tabcolsep}{1.0pc}
\vspace{1mm}
\begin{tabular}{c | c | c | c | c | c}
\specialrule{.15em}{.0em}{.15em}
\hline
\multicolumn{6}{c}{Comparison of the eigenvalues of the free scalar field theory with}\\
\multicolumn{6}{c}{the increasing resolution}\\
\hline
Analytical & $k=0$ & $k=1$ & $k=2$ & $k=3$ & $k=4$ \\
\hline
$0.000000$  & $0.000000$  & $0.000000$  & $0.000000$  & $0.000000$ & $0.000000$\\
$1.000000$  & $1.000000$  & $1.000000$  & $1.000000$  & $1.000000$ & $1.000000$\\
$2.000000$  & $2.000000$  & $2.000000$  & $2.000000$  & $2.000000$ & $2.000000$\\
$2.362020$  & $2.368770$  & $2.362481$  & $2.362049$  & $2.362021$ & $2.362020$\\
$3.000000$  & $3.000000$  & $3.000000$  & $3.000000$  & $3.000000$ & $3.000000$\\
$3.211938$  & $3.427144$  & $3.231760$  & $3.213295$  & $3.212025$ & $3.211944$\\
$3.362020$  & $3.368770$  & $3.362481$  & $3.362049$  & $3.362021$ & $3.362020$\\
$3.967989$  & $4.082342$  & $3.978361$  & $3.968697$  & $3.968034$ & $3.967992$\\
$4.000000$  & $4.000000$  & $4.000000$  & $4.000000$  & $4.000000$ & $4.000000$\\
\hline
\specialrule{.15em}{.15em}{.0em}
\end{tabular}
\end{center}
\end{table*}

The results presented in this section demonstrate that the low-energy spectrum of a quantum field theory can be computed accurately using only the effective scaling degrees of freedom obtained via the flow-equation method. The systematic convergence of eigenvalues with increasing resolution confirms the consistency of the approach and underscores the effectiveness of combining wavelet-based representations with similarity renormalization group techniques. This establishes a computationally efficient framework for extracting low-energy observables in quantum field theories. Extending and testing this method in interacting theories would be a natural next step.

\section{Summary and outlook}
\label{sec:conclusion_and_outlook}
In this work, we propose an effective Hamiltonian formulation of quantum field theories using the Daubechies wavelet basis. By employing the similarity renormalization group, we obtain an effective Hamiltonian that can be studied at significantly reduced computational cost. We demonstrate the method for the case of the 1+1-dimensional free scalar field theory. By implementing finite volume and resolution cutoffs, the Hamiltonian is represented in terms of a finite number of scaling and wavelet field operators and their conjugate momenta. At this stage, the flow-equation method of the similarity renormalization group is applied to decouple the scaling and wavelet degrees of freedom, yielding effective scaling and wavelet modes. To obtain the low-lying eigenvalues, we construct the Fock space basis using only the effective scaling degrees of freedom, which significantly reduces the dimension of the Hamiltonian matrix. The dimensions of the Hamiltonian matrix is further reduced by projecting out the zero momentum and even spatial parity sector of the Hamiltonian. The resulting matrix is then diagonalized to compute the low-lying eigenvalues of the theory for increasing resolution. The eigenvalues thus obtained approach the exact values, indicating that the method is consistent. It is noteworthy that, the dimensionality of the low-lying sector of the Hamiltonian can further be reduced by taking the projection of the Hamiltonian to the positive $\mathbb{Z}_2$ symmetric sector.

 The numerical results demonstrate that the low-lying eigenvalues systematically approach to their analytical values with increasing resolution, confirming the consistency of the method. The present formulation provides a systematic framework for separating scales in wavelet-based quantum field theories and offers a promising pathway for extending these methods to interacting field theories.

 An important extension of this work is application of the proposed framework to interacting quantum field theories, such as $\phi^4$ theory, where scale decoupling is expected to significantly reduce computational complexity. It would also be of interest to explore higher-dimensional theories using tensor-product wavelet bases, potentially establishing an efficient multiscale Hamiltonian approach for nonperturbative QFT calculations. Another possible extension the implementation of this framework on quantum computing platforms. The wavelet-based multiscale decomposition, combined with similarity renormalization group methods, yields a reduced effective Hamiltonian well suited for quantum simulation. This approach could enable studies of real-time dynamics and scattering processes in interacting quantum field theories, which remain difficult to access using classical methods.


\section*{Acknowledgement}
This work is supported by the Department of Atomic Energy, Government of India, under Project Identification Number RTI 4012. The simulations were performed on the computing clusters of the Department of Theoretical Physics, TIFR. MB acknowledges support from the visitor program of the Department of Theoretical Physics, TIFR.

\appendix
\section{Consistency of the wavelet-based creation and annihilation operators}
\label{appen:remarks_on_the_creation_and_annihilation_operator_construction}
The construction of creation and annihilation operators proposed in Ref.~\cite{PhysRevD.87.116011} is examined in this section and the key issues associated with this approach are highlighted.

As reported by Bulut et al.~\cite{PhysRevD.87.116011}, the expansion of field operators in terms of wavelet modes are given by,
\begin{eqnarray}
\phi(x) &=& 
\sum \frac{1}{\sqrt{2\gamma^{s,k}}}\, s^{k}_{m}(x)\left(a^{s,k}_m + a^{s,k\dagger}_m\right) \nonumber \\
&& +
\sum \frac{1}{\sqrt{2\gamma^{w,l}}}\, w^{l}_{m}(x)
\left(a^{w,l}_m + a^{w,l\dagger}_m\right),\\
\pi(x) &=& 
i \sum \sqrt{\frac{\gamma^{s,k}}{2}}\, s^{k}_{m}(x)\left(a^{s,k\dagger}_m - a^{s,k}_m\right) \nonumber \\
&& +
i \sum \sqrt{\frac{\gamma^{w,l}}{2}}\, w^{l}_{m,\alpha}(x)
\left(a^{w,l\dagger}_m - a^{w,l}_m\right),
\end{eqnarray}
where,
\begin{eqnarray}
a^{s,k}_n&:=& \frac{1}{\sqrt{2}}
\left( \sqrt{\gamma^{s,k}}\,\phi^{s,k}_n + i\,\frac{1}{\sqrt{\gamma^{s,k}}}\,\pi^{s,k}_n \right),\\
a^{w,r}_n &:=& \frac{1}{\sqrt{2}}
\left( \sqrt{\gamma^{w,r}}\,\phi^{s,k}_n+ i\,\frac{1}{\sqrt{\gamma^{w,r}}}\,\pi^{s,k}_n \right),
\end{eqnarray}
and,
\begin{eqnarray}
\label{eq:scaling_gamma}
\gamma^{s,k} &=& 
\frac{1 \pm \sqrt{1 - 4\bra{0} \phi^{s,k}_0\phi^{s,k}_0 \ket{0}
      \bra{0} \pi^{s,k}_0\pi^{k}_0 \ket{0}}}
     {2 \bra{0} \phi^{s,k}_0\phi^{s,k}_0 \ket{0}}, \\
\label{eq:wavelet_gamma}
\gamma^{w,r} &=&
\frac{1 \pm \sqrt{1 - 4\bra{0} \phi^{w,r}_0\phi^{w,r}_0 \ket{0}
      \bra{0} \pi^{w,r}_0\pi^{w,r}_0 \ket{0}}}
     {2 \bra{0} \phi^{w,r}_0\phi^{w,r}_0 \ket{0}},\nonumber\\
\end{eqnarray}
with,
\begin{gather}
\label{eq:scaling_phi_phi_inner_product}
\bra{0}\phi^{s,k}_0\phi^{s,k}_0 \ket{0}
= \frac{1}{2\pi}
\int \frac{\tilde{s}^{k*}_{0}(p)\tilde{s}^{k}_{0}(p)}{2\sqrt{p^2+1}} dp, \\
\label{eq:scaling_pi_pi_inner_product}
\bra{0} \pi^{s,k}_0\pi^{s,k}_0 \ket{0} =
\frac{1}{2\pi}
\int \frac{\tilde{s}^{k*}_{0}(p)\tilde{s}^{k}_{0}(p)\sqrt{p^2+1}}{2} dp,\\
\label{eq:wavelet_phi_phi_inner_product}
\bra{0} \phi^{w,r}_0\phi^{w,r}_0 \ket{0}
= \frac{1}{2\pi}
\int \frac{\tilde{w}^{r*}_{0}(p)\tilde{w}^{r}_{0}(p)}{2\sqrt{p^2+1}} dp, \\
\label{eq:wavelet_phi_phi_inner_product}
\bra{0} \pi^{w,r}_0\pi^{w,r}_0 \ket{0} =
\frac{1}{2\pi}
\int \frac{\tilde{w}^{r*}_{0}(p)\tilde{w}^{r}_{0}(p)\sqrt{p^2+1}}{2} dp,
\end{gather}    
and,
\begin{gather}
\tilde{s}^{k}_{0}(p)=\int s^{k}_0(x) e^{ipx}dx,\,\,\tilde{w}^{r}_{0}(p)=\int w^{r}_0(x)e^{ipx}dx.
\end{gather}
The field operator is real, i.e. $\phi^*(x)=\phi(x)$, which implies the coefficients, $\gamma^{s,k}$ and $\gamma^{w,r}$, appearing in the creation and annihilation operator are positive real numbers. However, their numerical evaluation shows that these coefficients are complex numbers, which is a clear contradiction with the above formulation. Moreover, even without numerical evaluation, we can show analytically that these quantities are in fact complex. The proof is presented below.

From the Cauchy-Schwarz inequality, from Eq.~(\ref{eq:scaling_phi_phi_inner_product}) and~(\ref{eq:scaling_pi_pi_inner_product}), we can write,
\begin{gather}
\label{eq:cauchy_schwarz_inequality}
\bra{0}\phi^{s,k}_0 \phi^{s,k}_0\ket{0}\bra{0}\pi^{s,k}_0 \pi^{s,k}_0\ket{0}\geq \left|\frac{1}{2\times2\pi}\int |\tilde{s}^k_n(p)|^2dp\right|^2,
\end{gather}
again, according to Parseval's theorem,
\begin{gather}
\label{eq:normalized_scaling_function}
\int |s^k_n(x)|^2 dx=\frac{1}{2\pi}\int |\tilde{s}^k_n(x)|^2=1.
\end{gather}
Substituting Eq.~(\ref{eq:normalized_scaling_function}) into~(\ref{eq:cauchy_schwarz_inequality}), we obtain,
\begin{eqnarray}
\bra{0}\phi^{s,k}_0 \phi^{s,k}_0\ket{0}\bra{0}\pi^{s,k}_0 \pi^{s,k}_0\ket{0}\geq \frac{1}{4}.
\end{eqnarray}
Similarly,
\begin{eqnarray}
\bra{0}\phi^{w,r}_0 \phi^{w,r}_0\ket{0}\bra{0}\pi^{w,r}_0 \pi^{w,r}_0\ket{0}\geq \frac{1}{4}.
\end{eqnarray}
These inequalities implies that the values of the constant coefficients, $\gamma^{s,k}$ and $\gamma^{w,r}$, given in Eq.~(\ref{eq:scaling_gamma}) and~(\ref{eq:wavelet_gamma}), respectively, are complex numbers. This observation motivates the development of an alternative wavelet-based formulation.

\section{The matrix plots of the evolution of $\Omega^2$ matrix with the increasing values of $\lambda$ for resolutions $1$, $3$, and $4$}
\label{appen:the_matrix_plots}
In this section, we present the remaining matrix plots of the SRG evolution of the matrix $\Omega^2$ with the increasing values of $\lambda$. In Figs.~\ref{fig:hamiltonian_increasing_lambda_res_1}, ~\ref{fig:hamiltonian_increasing_lambda_res_3}, and ~\ref{fig:hamiltonian_increasing_lambda_res_4} we show the matrix plots of the evolution of $\Omega^2$ for resolutions $1$, $3$, and $4$, respectively.
\begin{widetext}
\begin{figure*}[htbp]
    \centering
    \caption{\label{fig:hamiltonian_increasing_lambda_res_1}The matrix plot of the effective Hamiltonian of resolution $1$ with the increasing values of flowing parameter $\lambda$}
    \includegraphics[scale=0.35]{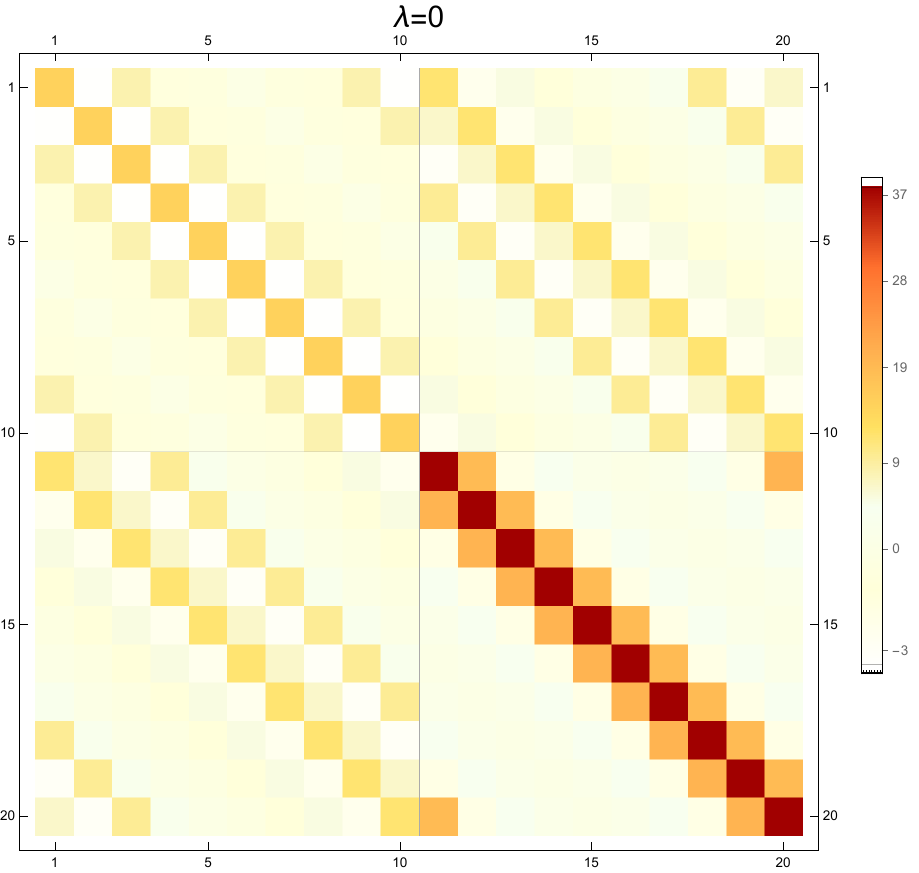}
    \includegraphics[scale=0.35]{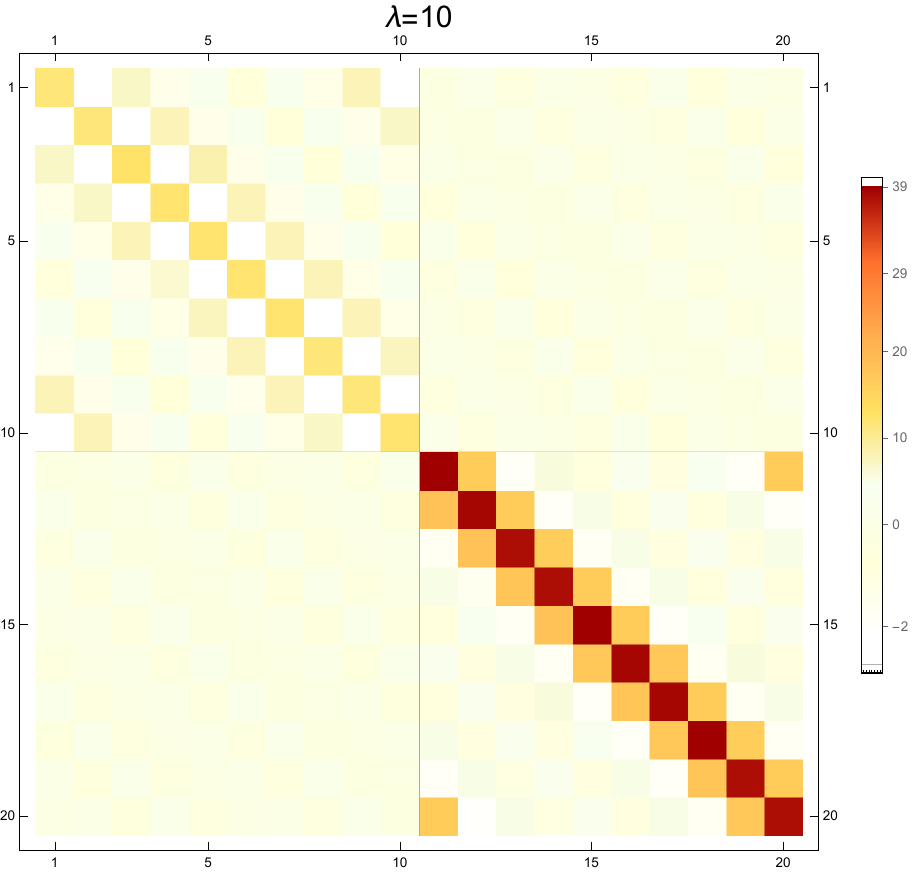}
    \includegraphics[scale=0.35]{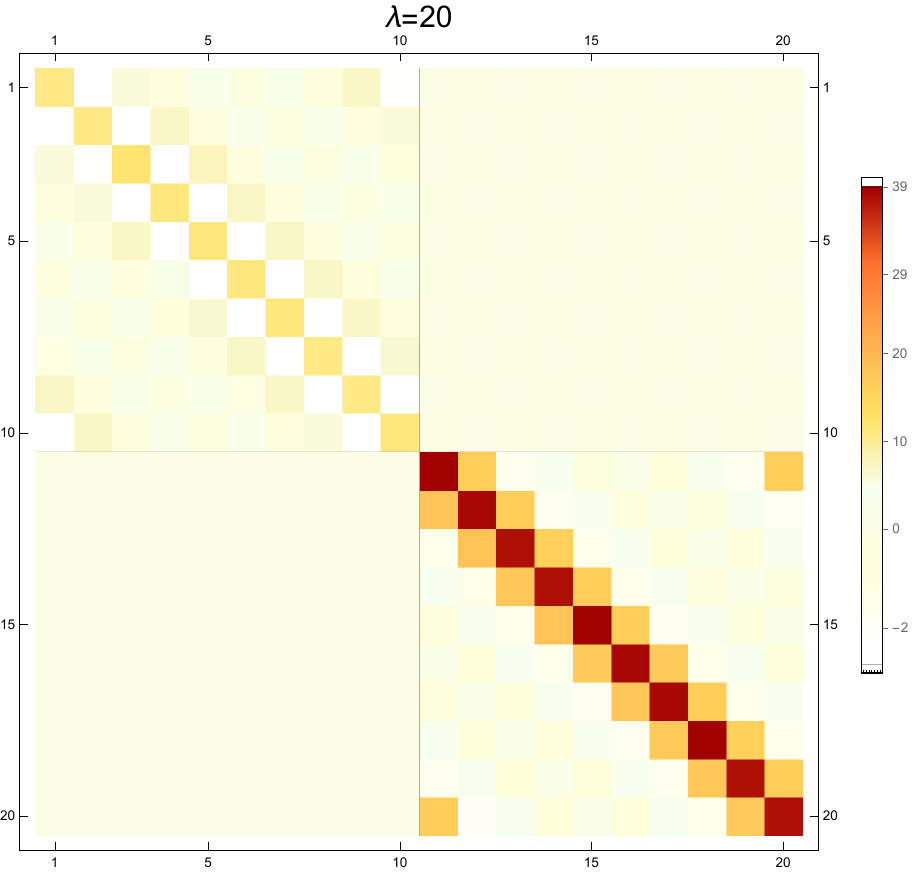}
\end{figure*}
\begin{figure*}[htbp]
    \centering
    \caption{\label{fig:hamiltonian_increasing_lambda_res_3}The matrix plot of the effective Hamiltonian of resolution $3$ with the increasing values of flowing parameter $\lambda$}
    \includegraphics[scale=0.3267]{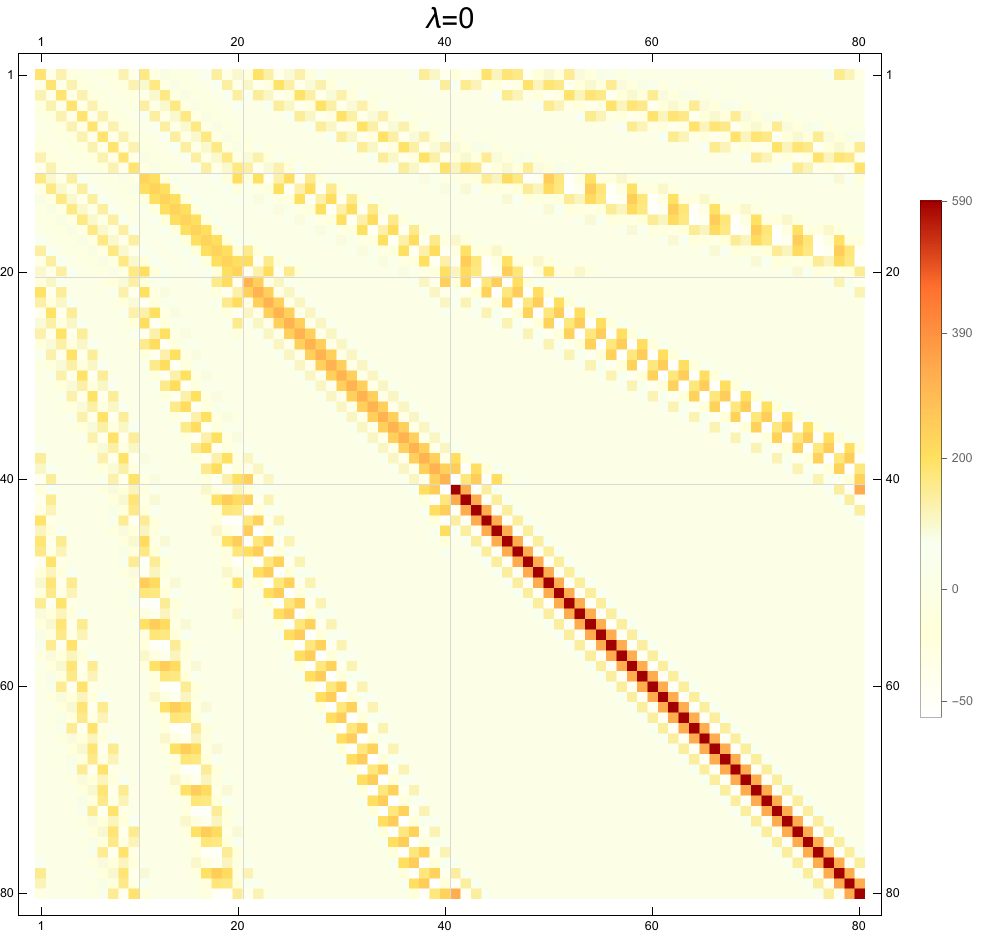}
    \includegraphics[scale=0.3267]{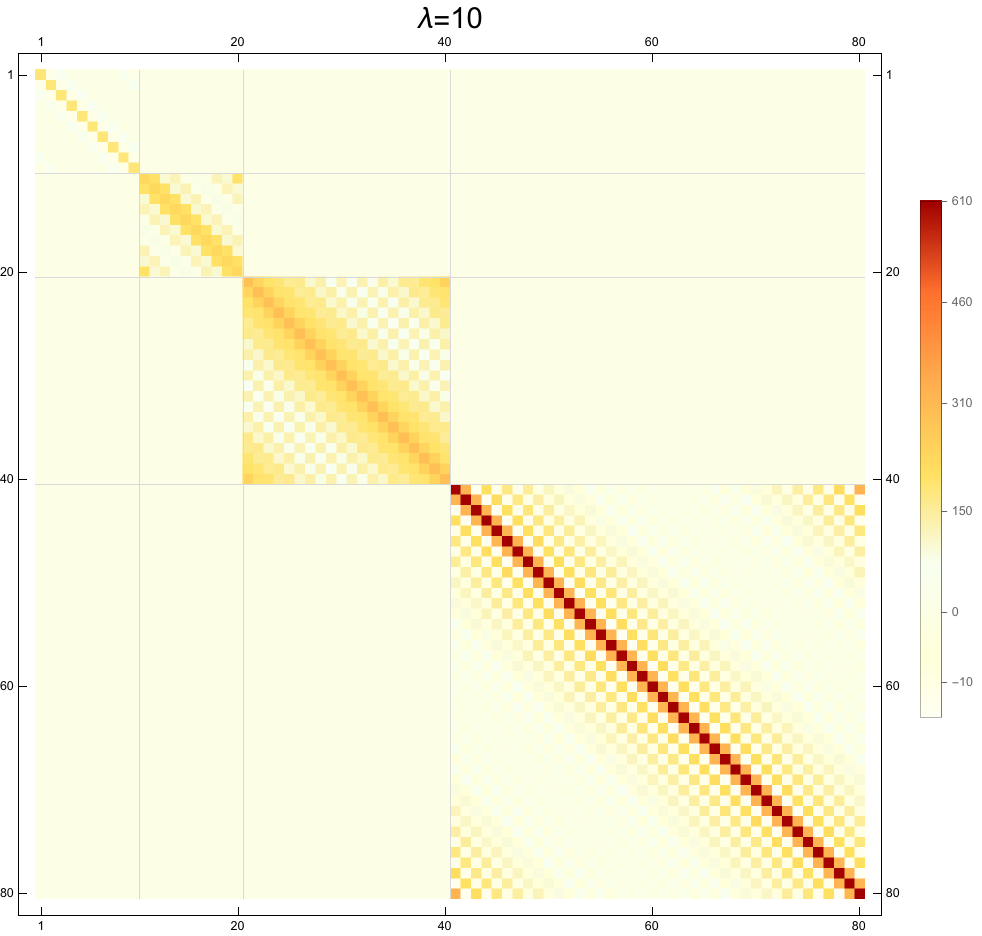}
    \includegraphics[scale=0.3267]{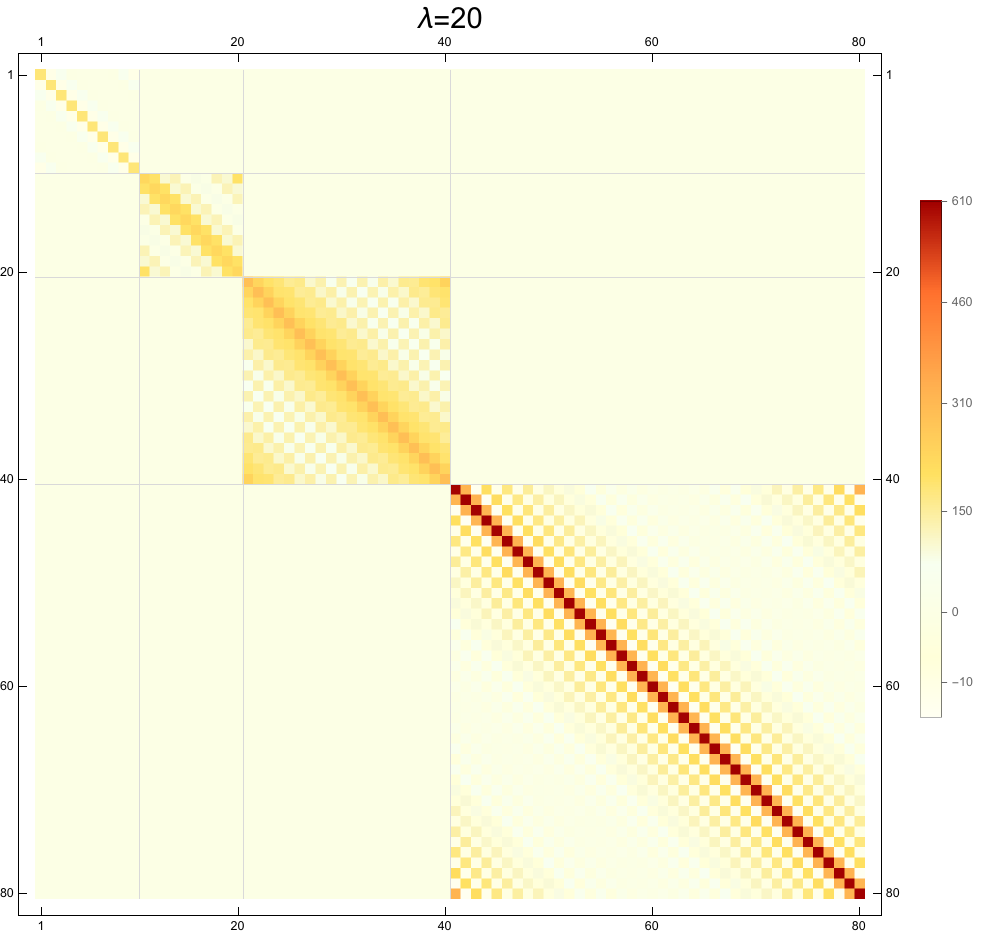}
\end{figure*}
\begin{figure*}[htbp]
    \centering
    \caption{\label{fig:hamiltonian_increasing_lambda_res_4}The matrix plot of the effective Hamiltonian of resolution $4$ with the increasing values of flowing parameter $\lambda$}
    \includegraphics[scale=0.3267]{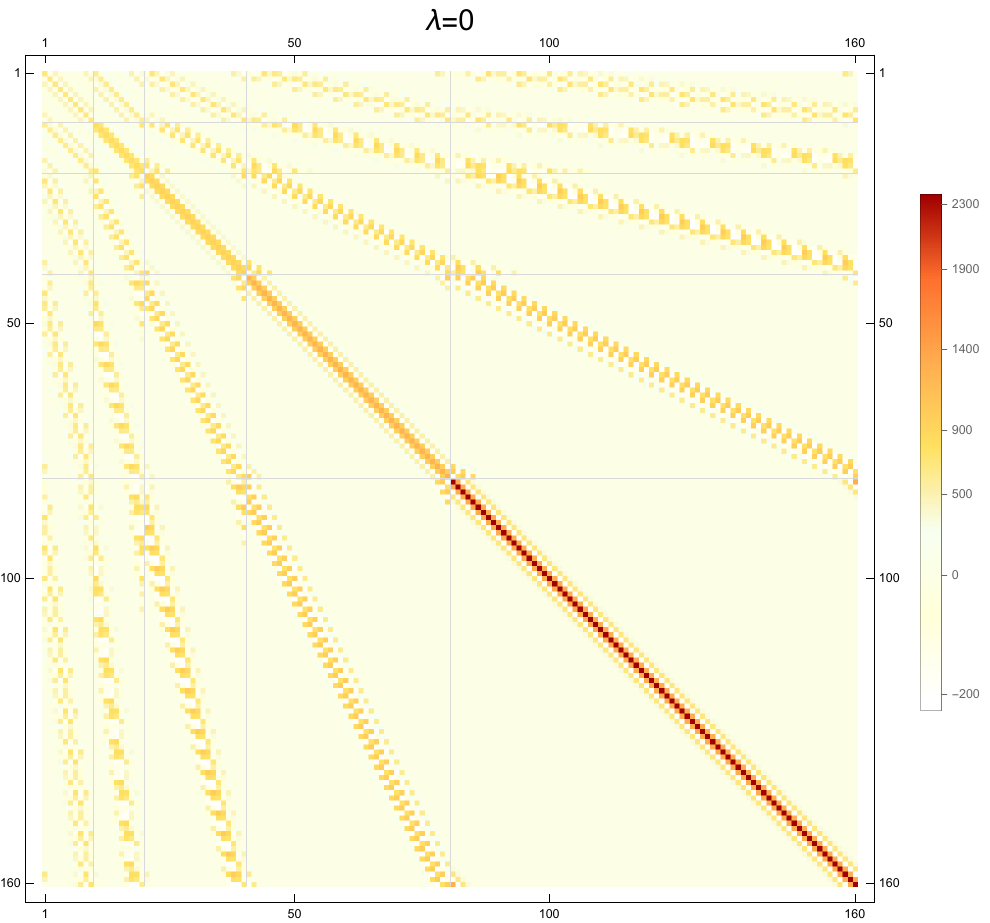}
    \includegraphics[scale=0.3267]{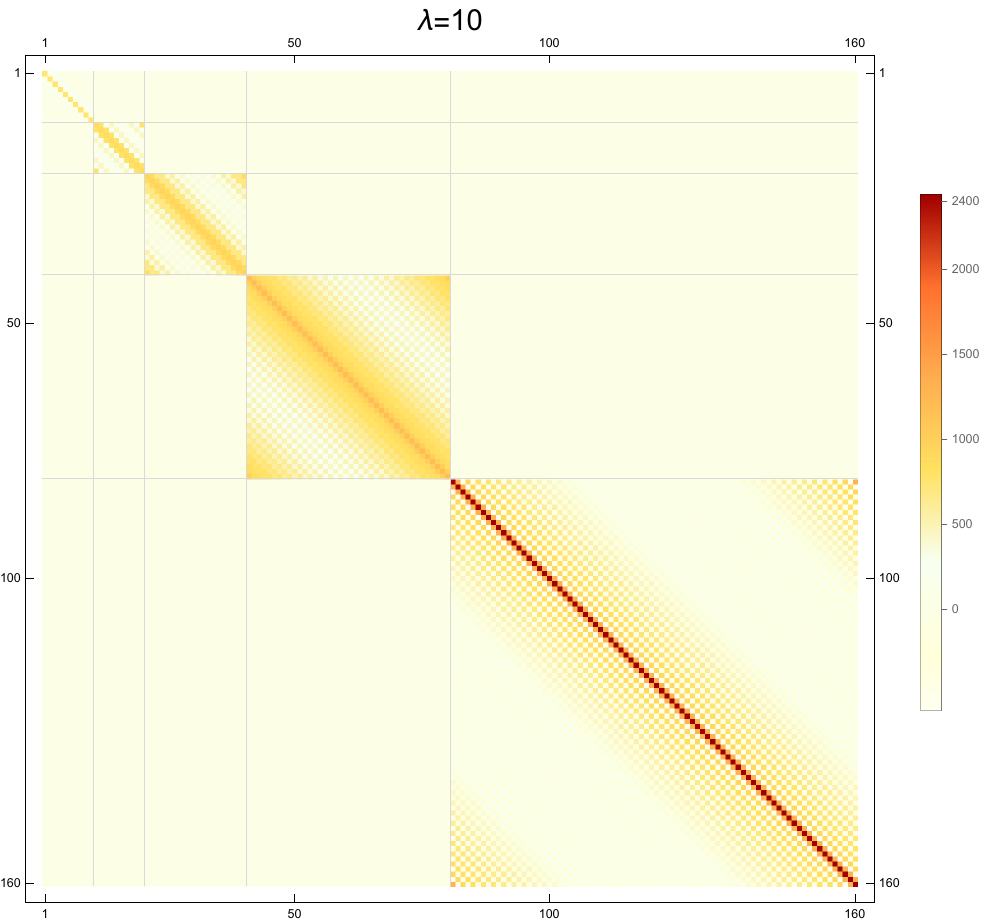}
    \includegraphics[scale=0.3267]{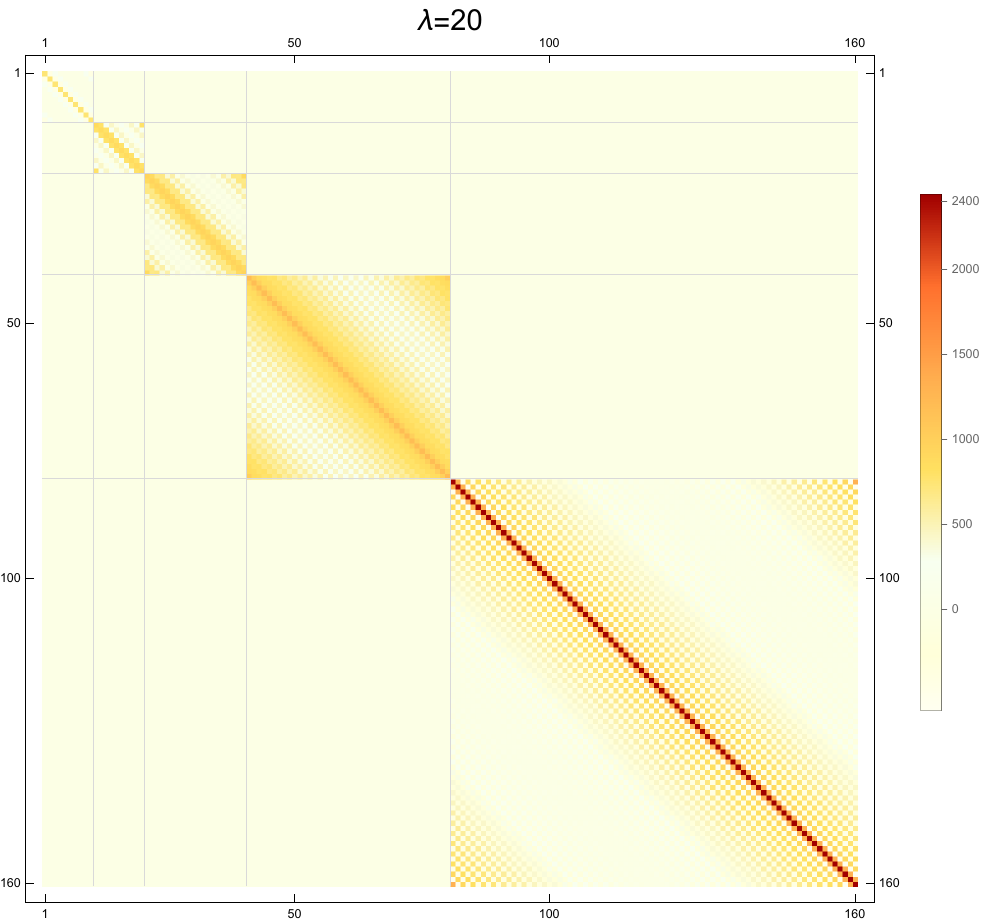}
\end{figure*}
\end{widetext}

\newpage

\section{Analytical evaluation of the overlap integrals}
\label{appen:overlap_integrals}
In this section of the appendix, we illustrate the detail procedure of analytically calculating the overlap integrals given in Eqs.~(\ref{eq:overlap_integral_derivative_scaling_scaling}) to~(\ref{eq:overlap_integral_derivative_scaling_wavelet}). 

We begin by writing the following properties of the translation ($\hat{T}$) and the dilation ($\hat{D}$) operators,
\begin{gather}
\label{eq:commutaion_of_D_T}
\hat{D}\hat{T}^{2k} = \hat{T}^{k}\hat{D}, \\
\label{eq:commutation_of_derivative_and_D}
\frac{d}{dx}\hat{D} = 2\hat{D}\frac{d}{dx}, \\
\label{eq:commutaion_of_x_and_D}
\hat{D}x = 2x\hat{D}, \\
\label{eq:commutaion_of_T_and_x}
\hat{T}x = (x-1)\hat{T} .
\end{gather}
In addition to these properties, we use the scaling equation and the definition of the mother wavelet and the derivatives of these equations,
\begin{eqnarray}
    s_m^{k}(x) &=& \sum_{n} H_{mn}\, s_n^{k+1}(x), \\
    w_m^{k}(x) &=& \sum_{n} G_{mn}\, s_n^{k+1}(x), \\
    s_{m}^{k\,\prime} &=& 2 \sum_{n} H_{mn}\, s_{n}^{(k+1)\,\prime}, \\
    w_{m}^{k\,\prime} &=& 2 \sum_{n} G_{mn}\, s_{n}^{(k+1)\,\prime},
\end{eqnarray}
where,
\begin{gather}
\begin{gathered}
    H_{mn}=h_{n-2m},\\
    G_{mn}=g_{n-2m}=(-1)^n h_{2K-1-(n-2m)}.
\end{gathered}
\end{gather}
By utilizing the properties of the dilation operator, $\hat{D}$, given in Eq.~(\ref{eq:commutation_of_derivative_and_D}), and above equations, the expression of the overlap integrals, $\mathcal{D}^k_{ss,mn}$, $\mathcal{D}^{lj}_{ww,mn}$, and $\mathcal{D}^{kl}_{sw,mn}$, given in Eqs.~(\ref{eq:overlap_integral_derivative_scaling_scaling}) to~(\ref{eq:overlap_integral_derivative_wavelet_wavelet}), can be rewritten as,
\begin{gather}
\begin{gathered}
    \mathcal{D}^k_{ss,mn}=\int s^{k\prime}_m(x) s^{k\prime}_n(x) dx \\
    = 2^{2k} \int s^{\prime}_m(x) s^{\prime}_n(x) dx =2^{2k}\mathcal{D}^0_{ss,mn},
\end{gathered}\\
\begin{gathered}
    \mathcal{D}^{lj}_{ss,mn} = \int w^{l\prime}_m(x) w^{j\prime}_n(x) dx \\
    = 2^{2(j+1)}\sum_{m^{\prime}n^{\prime}} \left(GH^{j-l}\right)_{mm^{\prime}}G_{nn^{\prime}} D^0_{ss,m^{\prime}n^{\prime}}, \,\, (j\geq l),
\end{gathered}\\
\begin{gathered}
    \mathcal{D}^{kl}_{ss,mn} = \int s^{k\prime}_m(x) w^{l\prime}_n(x) dx \\
    = 2^{2(l+1)}\sum_{m^{\prime}n^{\prime}} H^{l+1-k}_{mm^{\prime}}G_{nn^{\prime}} D^0_{ss,m^{\prime}n^{\prime}}, \,\, (l\geq k).
\end{gathered}
\end{gather}
As seen from the above expressions, each integral can be expressed in terms of $\mathcal{D}^{0}_{ss,mn}$,
\begin{eqnarray}
    \mathcal{D}^{0}_{ss,mn}=\int s^{\prime}_m(x) s^{\prime}_n(x)dx.
\end{eqnarray}
The translational invariant property of the overlap integrals further leads to,
\begin{eqnarray}
    \mathcal{D}^{0}_{ss,mn}=\int s^{\prime}_0(x) s^{\prime}_{n-m}(x)dx=\mathcal{D}^0_{ss,0p},
\end{eqnarray}
where, $p=n-m$.

\noindent Since the scaling function of order $K$ is non-zero over a spatial width $2K-1$, the value of the overlap integral is non-zero for a finite range of integer values of the index $p$ varying within $\left[-2K+2,2K-2\right]$. Hence, there exists a total $4K-3$ number of non-zero $\mathcal{D}^0_{ss,0p}$ for order $K$ Daubechies scaling functions. Subsequently, from the properties of the scaling functions, a set of homogeneous and inhomogeneous equations can be derived for the variables $\mathcal{D}^0_{ss,0p}$, which are to be solved to determine the overlap integrals.

To obtain the set of homogeneous equation, we start with the normalization condition of the scaling function,
\begin{eqnarray}
    \int s(x)dx=1.
\end{eqnarray}
Differentiating both side of the equation with respect to $x$, we get,
\begin{eqnarray}
\label{eq:normalization_condition_s_prime}
    \int s^{\prime}(x)dx=0.
\end{eqnarray}
Now, on the left-hand side of the above equation, forming a term of the type $s'(x)s'(x)$ requires a partition of unity in terms of $s'(x)$. To derive that, we first expand $x^n$ as a liner combinations of scaling functions as follows,
\begin{eqnarray}
     x^n=\sum_{l} b_l s_l(x).
\end{eqnarray}
To calculate the expansion coefficient $b_l$, we start with the definition of $b_l$,
\begin{gather}
    b_l = \int x^n s_l(x)dx,
\end{gather}
now, by using the unitarity of $\hat{D}$ and the scaling equation, from the above equation, we can derive,
\begin{eqnarray}
    b_l &=& \int x^n s(x - l) dx \nonumber \\
    &=&   \int (x+l)^n s(x) dx \nonumber \\
    \label{eq:the_expansion_coefficient_of_polynomial}
    &=& \sum_{q=0}^{n}\frac{n!}{q!(n-q)!} l^{n-q} \expval{x^n}_s,
\end{eqnarray}
where, $\expval{x^n}_s = \int x^n s(x) dx$, is called the moment of the scaling function. To evaluate this, we utilize the unitary property of the scaling function to rewrite the expression of the moment as,
\begin{gather}
    \expval{x^n}_s = \int \hat{D}^{-1} x^n \hat{D}^{-1} s(x) dx \nonumber \\
    = \frac{1}{2^n\sqrt{2}}\sum_r h_r \int x^n s(x-r) dx \nonumber \\
    = \frac{1}{2^n\sqrt{2}}\sum_r h_r \int (x+r)^n s(x) dx \nonumber \\
    =  \frac{1}{2^n\sqrt{2}}\sum_r h_r \sum_{u=0}^n \frac{n!}{u!(n-u)!}r^{n-u}\expval{x^n}_s,
\end{gather}
next, using $\sum_r h_r =\sqrt{2}$ and relocating the $u=n$ term to the left hand side of the above equation, we obtain the following recursion relation of $\expval{x^n}_s$:
\begin{gather}
    \expval{x^n}_s=\frac{1}{2^n - 1}\sum_{u=0}^{n-1}\frac{n!}{u!(n-u)!}\left(\sum_{r=1}^{2K-1}h_r r^{n-u}\right)\expval{x^r}_s.
\end{gather}
By recursively using the above expression, we can derive the moment of the scaling function, $\expval{x^n}_s$, for any arbitrary value of $n$, hence, the expansion coefficient $b_l$ for any polynomial of order, $n$, is obtained by substituting $\expval{x^n}_s$ in Eq.~(\ref{eq:the_expansion_coefficient_of_polynomial}).

By following the procedure described above, we can obtain the expansion coefficient $b_l$ for the expansion of $x$ in terms of the scaling function, $s(x)$:
\begin{eqnarray}
    b_l = l + \frac{1}{\sqrt{2}} \sum_{r}r h_r,
\end{eqnarray}
and, subsequently $x$ can be written as,
\begin{eqnarray}
    x = \sum_l l s_l(x) + \frac{1}{\sqrt{2}}\sum_r r h_r.
\end{eqnarray}
Now, taking derivative on both side of the above equation with respect to $x$:
\begin{eqnarray}
\label{eq:partition_of_unity_s_prime}
    \sum_l l s^{\prime}_l(x) = 1 .
\end{eqnarray}
Combining Eqs.~(\ref{eq:normalization_condition_s_prime}) and~(\ref{eq:partition_of_unity_s_prime}), we get the set of homogeneous equations associated with the coefficients $\mathcal{D}^0_{ss,0l}$:
\begin{eqnarray}
    \int \sum_l l s^{\prime}_l(x) s^{\prime}(x) dx &=& 0 \nonumber \\
    \label{eq:homogeneous_equation_of_D}
   \sum_l l \mathcal{D}^0_{ss,0l} &=& 0.
\end{eqnarray}

 To derive the inhomogeneous equation of the coefficient variables, $\mathcal{D}^0_{ss,0l}$, we begin by expanding $x^2$ in terms of $s(x)$, which is obtained by following the same procedure detailed above:
 \begin{eqnarray}
     x^2 = \sum_l \left(l^2 +2l \expval{x}_s + \expval{x^2}_s\right) s_l(x).
 \end{eqnarray}
 Taking derivative on both the sides of the above equation with respect to $x$, we get,
 \begin{eqnarray}
     2x = \sum_l \left(l^2 + 2l \expval{x}\right)s^{\prime}_l(x).
 \end{eqnarray}
 By utilizing the partition of unity property of Eq.~(\ref{eq:partition_of_unity_s_prime}), the above equation can be rewritten as,
 \begin{eqnarray}
     2x = \sum_l l^2 s^{\prime}_l(x) + 2 \expval{x}.
 \end{eqnarray}
 Next, multiplying both side of the above equation with $s^{\prime}(x)$ and integrating, we arrive at:
 \begin{gather}
     2\int x s^{\prime}(x) dx = \sum_l l^2 \int s^{\prime}_l(x)s^{\prime}(x)dx + 2 \expval{x} \int s^{\prime}(x) dx.
 \end{gather}
 Performing the partial derivative on the left hand side of the above equation and substituting the value of $\int s^{\prime}(x) dx= 0$ from Eq.~(\ref{eq:normalization_condition_s_prime}), we obtain the following equation:
 \begin{eqnarray}
     -2 &=& \sum_l l^2 \int s^{\prime}(x)s^{\prime}_l(x) dx\nonumber \\
  \implies -2 &=&\sum_{l} l^2 \mathcal{D}^0_{ss,0l}
 \end{eqnarray}
 This is the inhomogeneous equation, which along with the set homogeneous equations, Eq.~(\ref{eq:homogeneous_equation_of_D}), is solved to uniquely compute the values of the coefficients $\mathcal{D}^0_{ss,0l}$. In Table.~\ref{tab:the_overlap_integrals_of_D}, we presented the analytically calculated values of $\mathcal{D}^0_{ss,0l}$ for order $3$ ($K=3$) Daubechies scaling functions.
 \begin{table}[hbtp]
\begin{center}
\caption{\label{tab:the_overlap_integrals_of_D}The analytical values of the overlap integrals, $\mathcal{D}^0_{ss,0l}$}
\setlength{\tabcolsep}{2.0pc}
\vspace{1mm}
\begin{tabular}{c | c }
\specialrule{.15em}{.0em}{.15em}
\hline
Integral & Value \\
\hline
$\mathcal{D}^0_{ss,0(-4)}$  & $-3/560$ \\
$\mathcal{D}^0_{ss,0(-3)}$  & $-4/35$ \\
$\mathcal{D}^0_{ss,0(-2)}$  & $92/105$ \\
$\mathcal{D}^0_{ss,0(-1)}$  & $-356/105$ \\
$\mathcal{D}^0_{ss,00}$     &  $295/56$ \\
$\mathcal{D}^0_{ss,01}$     & $-356/105$ \\
$\mathcal{D}^0_{ss,02}$     & $92/105$ \\
$\mathcal{D}^0_{ss,03}$     & $-4/35$ \\
$\mathcal{D}^0_{ss,04}$     & $-3/560$ \\
\hline
\specialrule{.15em}{.15em}{.0em}
\end{tabular}
\end{center}
\end{table}

\section{Construction of the zero momentum sector}
\label{appen:zero_momentum}
Since the Fock-space basis elements in this formalism are not constructed from definite momentum modes, they are not eigenstates of the momentum operator. Instead, they are formed using position-space wavelet modes. However, momentum is the generator of spatial translations, with the translation operator given by $\hat{\mathbb{T}}(a)=e^{-\frac{i}{\hbar}a\hat{\mathrm{P}}}$. Therefore, it is possible to obtain eigenstates of the translation operator by forming suitable linear combinations of the Fock-space basis states. These combinations define sectors of definite momentum for the Hamiltonian. In particular, the zero-momentum sector is obtained by selecting the eigenstates of the translation operator with eigenvalue equal to unity. 

The steps to be followed for constructing translation-invariant states are:
\begin{enumerate}
    \item Generate all the raw Fock-space basis states within the resolution-truncated subspace of the Hilbert space.
    
    \item Identify the representative states among them. The representative state is the one that will be used to generate the translation-invariant state. A state is considered a representative state if it is the smallest (according to a chosen ordering) among all its translational copies.

    For example, consider the raw state $\ket{2,3,1,4}$. This is not the representative state because $2314$ is not the smallest number among all the translated copy of that state. The translated configurations are, $\ket{2,3,1,4}$, $\ket{4,2,3,1}$, $\ket{1,4,2,3}$, and $\ket{3,1,4,2}$. The smallest number is $1423$. Hence, The representative state of the state $\ket{2,3,1,4}$ is $\ket{1,4,2,3}$.

    \item After identifying the representative state, we form a linear combination of all the distinct translational copies of the state to construct a translation-invariant state that resides within the zero-momentum sector.

    For example, translationally invariant state $\ket{\phi}$ corresponding to the representative state $\ket{1,4,2,3}$ is,
    \begin{gather}
        \ket{\phi}\rightarrow \frac{1}{\sqrt{4}}\left(\ket{1,4,2,3}+\ket{3,1,4,2}+\ket{2,3,1,4}+\ket{4,2,3,1}\right).
    \end{gather}
    more generally, for a representative Fock basis state $\ket{n_1,n_2,\dots,n_N}$, a translationally invariant state is written as,
    \begin{gather}
        \ket{\phi_m}=\frac{1}{\sqrt{S_m}}\sum_{i=0}^{S_m-1} \hat{T}^i\ket{n_1,n_2,\dots,n_N},
    \end{gather}
    where, $T^i$ denotes the translation operator acting $i$ times, corresponding to a shift of the modes by $i$ lattice sites to the right, $S$ is the number of distinct translational copies. The value of $S$ is less than or equal to the number of modes $N$.

    \item Calculate the Hamiltonian matrix within this translation-invariant Fock-space basis, which yields the zero-momentum sector of the Hamiltonian in this framework:
    \begin{gather}
        \mathrm{H}_{mn}=\bra{\phi_m}\mathrm{H}\ket{\phi_n}\nonumber\\
        = \frac{1}{\sqrt{S_m S_n}}\sum_{i=0}^{S_m-1}\sum_{j=0}^{S_n-1}\bra{n_1,\dots,n_N}\hat{T}^{-i}\mathrm{H}\hat{T}^j\ket{n_1,\dots,n_N}\nonumber\\
        = \sqrt{\frac{S_m}{S_n}}\sum_{j=0}^{S_n-1}\bra{n_1,\dots,n_N}\mathrm{H}T^j\ket{n_1,\dots,n_N}.
    \end{gather}
\end{enumerate}

\section{Construction of the even and odd spatial parity sector}
\label{appen:parity_projection}
The action of the spacial parity operation on the Fock basis elements mirror the position space Fock basis. For example, the action of the parity operator on the Fock basis state, $\ket{n_1,n_2,\dots,n_{N-1},n_N}$ is given by,
\begin{eqnarray}
  \hat{\mathbb{P}} \ket{n_1,n_2\dots,n_{N-1},n_N}=\ket{n_N,n_{N-1},\dots,n_2,n_1}.
\end{eqnarray}
Therefore, the basis states fall into two categories.
\subsection{Symmetric eigenstates (Eigenvalues=$+1$)}
If the state is perfectly symmetric around the center of the lattice, the operation of the parity operator gives back the same state. For example, let's consider $\ket{1,0,3,0,1}$. The application of parity operator, $\hat{\mathbb{P}}$, onto this state returns the same state with eigenvalue $+1$.
\subsection{Asymmetric state (Not an eigenstate)}
In the wavelet based Fock space basis most of the states fall under this category. For example, let's consider a state $\ket{2,0,0,0,0}$. The parity operation inverses the state to $\ket{0,0,0,0,2}$. So, $\ket{2,0,0,0,0}$ is not an eigenstate of $\hat{\mathbb{P}}$. Although using these states, we can construct the eigen basses of even parity and the odd parity sector of the Hilbert space.

To do that, let's consider, an asymmetric state $\ket{A}$, whose flipped version is $\ket{B}$,
\begin{itemize}
\item Construction of the even parity state:

then, the even parity eigenstate is constructed by taking the linear combination of $\ket{A}$ and $\ket{B}$ as, $\ket{S_+}=\frac{1}{2}(\ket{A}+\ket{B})$. The application of the parity operator onto this state gives the eigenvalue $+1$:
\begin{gather}
   \hat{\mathbb{P}}\ket{S_+}= \frac{1}{2}\hat{\mathbb{P}}\left(\ket{A}+\ket{B}\right)=\frac{1}{2}\left(\ket{B}+\ket{A}\right)=(+1)\ket{S_+},
\end{gather}
\item Construction of the odd parity state:

and, the odd parity eigenstate is constructed by taking the other possible linear combinatin of $\ket{A}$ and $\ket{B}$ as, $\ket{S_-}=\frac{1}{2}(\ket{A}-\ket{B})$. The application of the parity operator onto this state returns the eigenvalue $-1$:
\begin{gather}
   \hat{\mathbb{P}}\ket{S_-}= \frac{1}{2}\hat{\mathbb{P}}\left(\ket{A}-\ket{B}\right)=\frac{1}{2}\left(\ket{B}-\ket{A}\right)=(-1)\ket{S_-}.
\end{gather}
\end{itemize}
\bibliography{References}
\end{document}